\documentclass[12pt]{article}
\pdfoutput=1

\usepackage{amsthm,amsfonts,amssymb,amsmath,dsfont}
\usepackage[pdftex]{graphicx,graphics,color}
\usepackage{color,colortbl}
\usepackage{lmodern}
\usepackage[english]{babel}
\definecolor{citecl}{rgb}{0.55,0.55,0.55}
\usepackage[colorlinks=true,linkcolor=black,citecolor=citecl,anchorcolor=blue,filecolor=blue,urlcolor=blue]{hyperref}
\usepackage[super,sort&compress]{natbib}
\usepackage[affil-it]{authblk}
\usepackage{subfigure}
\usepackage{hyphenat}

\def\R{\mathbb{R}}

\newcommand{\onlinecite}[1]{\hspace{-1 ex} \nocite{#1}\citenum{#1}} 

\begin{document}

% Declaração de comandos %%%%%%%%%%%%%%%%%%%%%%%%%%%%%%%%%%%%%%%%%%%%%%%%%%%%%%%%%%%%%%%%%%%%%%%%%%%%%%%%%%%%%%%%%%%%%%%%%%%%%%%%%%%%%%%%%%

\newcommand{\nn}{\nonumber}
\newcommand{\bra}{\langle}
\newcommand{\ket}{\rangle}
\newcommand{\del}{\partial}
\newcommand{\vt}{\vec}
\newcommand{\dg}{^{\dag}}
\newcommand{\cg}{^{*}}
\newcommand{\T}{^{T}}
\newcommand{\vep}{\varepsilon}
\newcommand{\vth}{\vartheta}
\newcommand{\suml}{\sum\limits}
\newcommand{\prodl}{\prod\limits}
\newcommand{\intl}{\int\limits}
\newcommand{\til}{\tilde}
\newcommand{\wb}{\overline}
\newcommand{\mcl}{\mathcal}
\newcommand{\mfk}{\mathfrak}
\newcommand{\mds}{\mathds}
\newcommand{\mbb}{\mathbb}
\newcommand{\mrm}{\mathrm}
\newcommand{\ds}{\displaystyle}
\newcommand{\rmi}{\mathrm{i}}
\newcommand{\rme}{\mathrm{e}}
\newcommand{\rmd}{\mathrm{d}}
\newcommand{\rmD}{\mathrm{D}}
\newcommand{\vphi}{\varphi}
\newcommand{\stm}{\text{\textsf{s}}}
\newcommand{\bth}{\text{\textsf{b}}}
\newcommand{\itr}{\text{\textsf{i}}}
\renewcommand{\mod}{\mathrm{\,mod\,}}
\renewcommand{\b}{\bar}
\renewcommand{\dim}{\mbox{dim}}
\renewcommand{\-}{\,-}

% Outras opções %%%%%%%%%%%%%%%%%%%%%%%%%%%%%%%%%%%%%%%%%%%%%%%%%%%%%%%%%%%%%%%%%%%%%%%%%%%%%%%%%%%%%%%%%%%%%%%%%%%%%%%%%%%%%%%%%%%%%%%%%%%

\interfootnotelinepenalty=10000

\renewcommand{\topfraction}{0.85}
\renewcommand{\textfraction}{0.1}

\allowdisplaybreaks[1]

\setlength{\jot}{1ex}

% Title %%%%%%%%%%%%%%%%%%%%%%%%%%%%%%%%%%%%%%%%%%%%%%%%%%%%%%%%%%%%%%%%%%%%%%%%%%%%%%%%%%%%%%%%%%%%%%%%%%%%%%%%%%%%%%%%%%%%%%%%%%%%%%%%%%%

\title{A method for Hamiltonian truncation: A four-wave example}

\author[1]{Thiago F. Viscondi}
\author[1]{Iber\^e L. Caldas}
\author[2]{Philip J. Morrison}
\affil[1]{Institute of Physics, University of S\~ao Paulo, S\~ao Paulo, SP, Brazil}
\affil[2]{Department of Physics and Institute for Fusion Studies, The University of Texas at Austin, Austin, TX 78712-1060, USA}

\date{}
\maketitle

% Abstract %%%%%%%%%%%%%%%%%%%%%%%%%%%%%%%%%%%%%%%%%%%%%%%%%%%%%%%%%%%%%%%%%%%%%%%%%%%%%%%%%%%%%%%%%%%%%%%%%%%%%%%%%%%%%%%%%%%%%%%%%%%%%%%%

A method for extracting  finite-dimensional Hamiltonian systems from a class of $2+1$ Hamiltonian mean field theories 
is presented. These theories possess noncanonical Poisson brackets, which normally resist Hamiltonian truncation, but 
a process of beatification by coordinate transformation near a reference state is described in order to perturbatively 
overcome this difficulty. Two examples of four-wave truncation of Euler's equation for scalar vortex dynamics are given 
and compared: one a direct non-Hamiltonian truncation of the equations of motion, the other obtained by beatifying the 
Poisson bracket and then truncating.

% Section1 %%%%%%%%%%%%%%%%%%%%%%%%%%%%%%%%%%%%%%%%%%%%%%%%%%%%%%%%%%%%%%%%%%%%%%%%%%%%%%%%%%%%%%%%%%%%%%%%%%%%%%%%%%%%%%%%%%%%%%%%%%%%%%%%
 
% Section1 %%%%%%%%%%%%%%%%%%%%%%%%%%%%%%%%%%%%%%%%%%%%%%%%%%%%%%%%%%%%%%%%%%%%%%%%%%%%%%%%%%%%%%%%%%%%%%%%%%%%%%%%%%%%%%%%%%%%%%%%%%%%%%%%
\section{Introduction}
\label{sec:1}

The reduction of partial differential equations describing physical phenomena, infinite\hyp{}dimensional dynamical systems,  
to ordinary differential equations, finite-dimensional dynamical systems, is a mainstay procedure of physics. This is 
done on the one hand in order to obtain semi-discrete schemes for numerical computation and on the other to obtain reduced 
low-dimensional nonlinear models for  describing  specific physical mechanisms. Examples of the former include finite 
difference methods such as the Arakawa Jacobian scheme (e.g.\ Refs.~\onlinecite{Arakawa66,Arakawa81}) and generalizations 
(e.g.\ Ref.~\onlinecite{Salmon89}) or the discontinuous Galerkin method (e.g.\ Refs.~\onlinecite{Cockburn91,Cheng13}), 
while examples of the latter are low-order modal models. The focus of the present paper is to describe a method for 
obtaining weakly nonlinear Hamiltonian models from noncanonical Hamiltonian systems,\cite{Morrison80,Morrison98} models 
that can then be truncated to obtain finite-dimensional Hamiltonian systems.  The method is described in general terms 
and demonstrated explicitly by extracting a four-wave model from Euler's equation for vorticity dynamics in two dimensions 
as an example. 

Although a plethora of low-order models have been obtained by various means, the three-wave model is an exact 
highly studied case that can be extracted  from physical systems that describe, e.g., fluids, plasmas, and optics 
(e.g.\ Refs.~\onlinecite{WW77,Wersinger80,Lopes96}).  Similarly,  four-wave models have been widely derived and 
studied (e.g.~Refs.~\onlinecite{Sugihara68,Karplyuk73,Turner80,Verheest82,Romeiras83,Chian96,Pakter97}). These 
reductions are obtained  from a parent model, a nonlinear partial differential equation, by linearizing about 
an equilibrium state and analyzing the dispersion relation for the possibilities of three or four-wave resonances 
between linear eigenmodes. If such exist, these models can be derived by averaging or other means. 

Of particular current interest are low-order models for describing the dynamics of zonal flows that occur in 
geophysical fluid dynamics and plasma physics, in the contexts of planetary atmospheres and the edge tokamaks, 
respectively. These separate fields of research have common physics as captured by the Charney-Hasegawa-Mima 
(or quasigeostrophic) equation,\cite{Charney71,Hasegawa78} which describes both Rossby waves and plasma drift 
waves (e.g.\ Ref.~\onlinecite{Connaughton10}).  In order to describe effects in tokamaks such as the transition 
to turbulence due to gradients, the emergence of zonal flows, and barriers to transport, four-wave\cite{LashmoreDavies01,
LashmoreDavies05} and higher dimensional models \cite{Kolesnikov05,Dewar07} have been proposed, but the Hamiltonian 
form has either not been determined or has not been obtained from a parent model. The methods of this paper provide 
a means for doing this. 

In general, when dissipative effects are ignored,  one may expect systems to possess Hamiltonian structure. This is 
the case for the three-wave model and is indeed overwhelming the case  for systems that describe fluids, plasmas, and 
other kinds of matter.\cite{Morrison98}  The Hamiltonian structure provides access to the great body of lore about such 
systems; e.g., it  is known at the outset that  only certain dispersion relations and  bifurcations are possible, the 
structure provides a means for determining nonlinear stability,  and  because of the well-known theorem of Liouville 
on  the incompressibility of phase space, attractors are not possible  (see e.g.\  Refs.~\onlinecite{Morrison05,Tassi08} 
for examples). 

For the three-wave model,   the Hamiltonian form  was  identified after its derivation; however, this form can be 
obtained directly from the noncanonical Hamiltonian structure, i.e., one in terms of a Poisson bracket in noncanonical 
variables (see Ref.~\onlinecite{Morrison98})  of the  parent model  (e.g.\ Refs.~\onlinecite{Kueny95a,Kueny95b}), in 
which case it is seen that the Manley-Rowe relations and other invariants are directly obtained from the Hamiltonian. 
Similarly, for the four-wave model that describes modulational instability of surface water waves,  the Hamiltonian 
structure can be obtained from  a Lagrangian, or equivalently canonical Hamiltonian, description  of a parent 
model (e.g.\ Ref.~\onlinecite{Zakharov09}).  We mention that sometimes Hamiltonian four-wave models are proposed 
a priori\cite{Rizzato15} and then the tools of Hamiltonian dynamics are exploited. 

Hamiltonian reduction of  the noncanonical Poisson brackets considered here, by means of direct projection onto  
bases such as Fourier series, is in general not possible because truncation destroys the Jacobi identity.\cite{Morrison81a}  
In the case of two dimensions, the procedure of Refs.~\onlinecite{Kueny95a,Kueny95b} that addressed a beam plasma 
system with one dimension  is not workable. This is because the two-dimensional Poisson bracket that describes, e.g., 
the Charney-Hasegawa-Mima equation and  Vlasov-Poisson system (see section~\ref{sec:2}) depends on the dynamical 
variable,\cite{Morrison80,Morrison82b} and this is the source of difficulty. 

In previous work\cite{MorrisonVanneste} this difficulty was surmounted by a process called \textit{beatification}, 
whereby the dynamical variable is removed from the noncanonical Poisson bracket to lowest order by a perturbative 
transformation about an equilibrium state. In removing the dynamical variable from the Poisson bracket, beatification 
is a step toward canonization, i.e., transforming to variables in which the Poisson bracket has standard canonical form.  
The beatification procedure, in eliminating variable dependence from the Poisson bracket, increases the degree of nonlinearity 
of the Hamiltonian. Then, the Poisson bracket and Hamiltonian can be expanded in a basis and then truncated with the Hamiltonian 
form being preserved. The  present paper builds on previous work\cite{MorrisonVanneste} by generalizing to expansion about an 
arbitrary reference state and by continuing the expansion to one higher order. Our example, which as noted above starts from 
Euler's equation, is similar but not equivalent to the four-wave model of Refs.~\onlinecite{LashmoreDavies01,LashmoreDavies05},  
since the latter starts from the modified Hasegawa-Mima equation.  

We note that there is literature on the general problem of extracting reduced noncanonical Hamiltonian systems from a 
parent Hamiltonian system by expansion in a small parameter. A procedure was introduced in Refs.~\onlinecite{Olver84a,Olver84b} 
with water waves as an application, described as a general deformation of Poisson brackets on Poisson manifolds in 
Ref.~\onlinecite{Vorobev88}, and placed in the context of generalized Lie transforms in Ref.~\onlinecite{FloresEspinoza13}. 
Central to all these developments is the Schouten bracktet\cite{Schouten53}. These works concern expansion about a dynamical 
system in order to produce a new, possibly noncanonical, Hamiltonian dynamical system, while beatification is an expansion 
about a phase space point that produces a constant Poisson bracket.

The paper is organized as follows. In section~\ref{sec:2}, we introduce a general class of Hamiltonian systems that 
share a common noncanonical Poisson bracket and associated Casimir invariants, constants of motion associated with 
the bracket degeneracy. Depending on the choice of Hamiltonian, this class includes the Hamiltonian descriptions of 
the Vlasov-Poisson system, quasigeostrophy and other mean field theories, but of main concern is the example we treat, 
the two-dimensional Euler system for the dynamics of the scalar vorticity. In section~\ref{sec:3} we perform a direct  
truncation. To this end the noncanonical Poisson bracket is transformed in subsection \ref{ssc:3.1} by considering 
dynamics relative to an arbitrary given reference state. The reference state for our four-wave example is introduced 
here. Then in subsection~\ref{ssc:3.2} the transformed Poisson bracket is re-expressed by expanding the new dynamical 
variable in terms of a Fourier series. With any Hamiltonian written in terms of the Fourier series, an infinite-dimensional 
dynamical system is obtained for the Fourier amplitudes. This is worked out for the Euler example. A truncation is done 
in subsection~\ref{ssc:3.3}, producing a four-wave system, which is seen in subsection~\ref{ssc:3.4} to conserve a reduced 
form of the energy and to possess  a remnant of a Casimir invariant of the unreduced system. However, this system is shown 
in subsection~\ref{ssc:3.5} not to be Hamiltonian because the  truncated bracket does not satisfy the Jacobi identity. 
In section~\ref{sec:4} we describe beatification, the procedure by which variable dependence is removed from the noncanonical 
Poisson bracket, and then we apply it to the bracket presented in section~\ref{sec:2}. Fourier expansion of the beatified 
Poisson bracket is done in subsection~\ref{ssc:5.1} which prepares the way for  Hamiltonian truncation. Although one 
can truncate by retaining any number of Fourier amplitudes, we demonstrate the method for our four-wave example in 
subsection~\ref{ssc:5.2}. Contrary to subsection~\ref{ssc:3.4}, it is observed in subsection~\ref{ssc:5.3} that two 
Casimir invariants are obtained for our four-wave Hamiltonian example. In section \ref{sec:6} we use the notion of 
a recurrence plot to give some preliminary numerical evidence for the superiority of the Hamiltonian truncation of 
subsection~\ref{ssc:5.2}. Finally, in section~\ref{sec:7}, we summarize the main findings of our work and make 
some concluding remarks.

%%%%%%%%%%%%%%%%%%%%%%%%%%%%%%%%%%%%%%%%%%%%%%%%%%%%%%%%%%%%%%%%%%%%%%%%%%%%%%%%%%%%%%%%%%%%%%%%%%%%%%%%%%%%%%%%%%%%%%%%%%%%%%%%%%%%%%%%%%%

% Section2 %%%%%%%%%%%%%%%%%%%%%%%%%%%%%%%%%%%%%%%%%%%%%%%%%%%%%%%%%%%%%%%%%%%%%%%%%%%%%%%%%%%%%%%%%%%%%%%%%%%%%%%%%%%%%%%%%%%%%%%%%%%%%%%%

% Section2 %%%%%%%%%%%%%%%%%%%%%%%%%%%%%%%%%%%%%%%%%%%%%%%%%%%%%%%%%%%%%%%%%%%%%%%%%%%%%%%%%%%%%%%%%%%%%%%%%%%%%%%%%%%%%%%%%%%%%%%%%%%%%%%%
\section{A class of Hamiltonian systems}
\label{sec:2}

We begin by describing a general class of Hamiltonian systems, 2+1 mean field theories.  
First we give the Hamiltonian then describe the Poisson bracket and associated Casimir 
invariants.  

We take as a basic dynamical variable a scalar density or vorticity-like quantity, $\omega(r,t)$, which is a 
real-valued function defined on a two-dimensional domain~$\mcl{D}$. For the present development we assume 
Cartesian coordinates where  $r=(x,y)\in \mcl{D}$. A general class of Hamiltonian mean field theories\cite{Morrison03} 
possess a Hamiltonian (energy) functional contained in the  following form:
\begin{equation}
 H[\omega]=  \int_{\mcl{D}}\rmd^{2}r\;\omega(r,t)\, h_1(r) 
 + \frac{1}{2}\int_{\mcl{D}}\rmd^{2}r\!\int_{\mcl{D}}\rmd^{2}r'\;\omega(r,t)\, h_2(r;r')\, \omega(r',t) \,,
 \label{eq:2.1}
\end{equation}
where $\rmd^{2}r=\rmd x \rmd y$. The  first term of \eqref{eq:2.1}, the inertial term, represents energy 
associated with free motion as determined by the function $h_1$, while the second term, the interaction 
term, represents the energy of two-point interaction as determined by the function $h_2$.  One could 
generalize this with three-point and higher interactions in an obvious way.  

Hamiltonians of the form of \eqref{eq:2.1} include the examples below for well-known systems: 
\begin{itemize}
\item when $\omega$ is the phase space density for a species of mass $m$ and charge $e$, $\mcl{D}=\R^2$ 
the phase space for a one degree-of-freedom system, for which $r=(x,v)\in \mcl{D}$ with velocity $v$, 
kinetic energy $h_1= mv^2/2$, and interaction potential $h_2=e|x-x'|$ for charged sheets, \eqref{eq:2.1} 
is the energy for the Vlasov-Poisson system.\cite{Morrison80b} 

\item when $\omega$ is the scalar vorticity, then $\mcl{D}$ denotes the planar domain occupied by the fluid.  
Upon choosing $h_1\equiv 0$ and defining $\Delta^{-1}$ to be the formal inverse of the two-dimensional Laplacian 
operator  $\Delta=\del_{x}^{2}+\del_{y}^{2}$, then \eqref{eq:2.1} is the Hamiltonian for Euler's equation 
describing an ideal, incompressible and two-dimensional fluid\cite{Morrison82b,Morrison98}  
\begin{equation}
 H[\omega]= -\frac{1}{2}\int_{\mcl{D}}\rmd^{2}r\;\omega \Delta^{-1} \omega\,.
 \label{euler}
\end{equation} 
For this case $h_2$ is proportional to the Green's function corresponding  with $\Delta$.  
This case will be the starting point for the four-wave example treated in our paper. 

\item when  $\omega$ is the charge density for drift waves or the potential  vorticity of geophysical 
fluid dynamics, then $\omega = b(x) -\mathcal{L}\psi$, where for the Hasegawa-Mima equation or quasigeostrophy 
$\mathcal{L}:=  \Delta + \kappa^2$,  $\psi$ is the electrostatic potential or stream function,  and $b$ represents 
the electron density or $\beta$-effect, respectively, with  $\kappa^{-1}$ measuring  the Rossby deformation radius. 
For this case the Hamiltonian is\cite{Tassi09}
\begin{equation}
 H[\omega]=   \int_{\mcl{D}}\rmd^{2}r\left(\omega\, \mathcal{L}^{-1}b  
 - \frac{1}{2}\, \omega \mathcal{L}^{-1} \omega\right) \,.
 \label{genHam}
\end{equation} 
Note, $\mathcal{L}$ could be any invertible elliptic operator. 
\end{itemize}

To define the Poisson bracket we require the functional derivative, which is defined as usual by  
\begin{equation}
\delta H[\omega; \delta \omega]
= \left.\frac{d}{d\epsilon} H[\omega + \epsilon \delta \omega]\right|_{\epsilon =0} 
= \int_{\mcl{D}}\rmd^{2}r\,  \delta \omega \frac{\delta H}{\delta \omega}\,, 
\end{equation}
where $\delta \omega(r,t)$ is a variation of $\omega$. 
(See e.g. Ref.~[\onlinecite{Morrison98}] for details.)
For the Hamiltonian of \eqref{eq:2.1} we have 
\begin{equation}
 \frac{\delta H}{\delta \omega}=  
 h_1 +  \int_{\mcl{D}}\rmd^{2}r'\;  h_2(r;r')\, \omega(r',t)  \,,
 \label{dHdom}
\end{equation}
a quantity that will be inserted into a  Poisson bracket. The Poisson bracket for our class 
of theories is given by the following bilinear product between two arbitrary functionals of 
the field~$\omega$:
\begin{equation}
 \{F,G\}=\int_{\mcl{D}}\rmd^{2}r\;\omega\left[\frac{\delta F}{\delta \omega},\frac{\delta G}{\delta \omega}\right]
 =\int_{\mcl{D}}\rmd^{2}r\;\frac{\delta F}{\delta \omega}\mcl{J}(\omega)\frac{\delta G}{\delta \omega}\,,
 \label{eq:2.2}
\end{equation}
where $[f,g]=(\del_{x}f)(\del_{y}g)-(\del_{y}f)(\del_{x}g)$, with $f$ and $g$ being two arbitrary functions 
on the domain~$\mcl{D}$. Proofs of the Jacobi identity for \eqref{eq:2.2} were given by direct computation 
in Ref.~\onlinecite{Morrison81a} and by Clebsch reduction in Ref.~\onlinecite{Morrison82b}. We note, it 
can also be shown by the vanishing of the Schouten bracket (e.g.\ Ref.~\onlinecite{Olver84a}). Comparison 
of the two integrals of \eqref{eq:2.2} gives the Poisson operator\footnote{This quantity has various names. 
For canonical systems it would naturally be called the cosymplectic operator because it is dual to the symplectic 
two-form. However, because it is degenerate, one could call it by the awkward  `copresymplectic' form! Another 
name, one we will use for finite-dimensional systems (cf.\ section~\ref{sec:4}), is the Poisson matrix.}
\begin{equation}
 \mcl{J}(\omega)f=-\left[\omega,f\right]\,,
 \label{eq:Popt}
\end{equation}
in which $f$ is again an arbitrary function. Notice that the two integrals shown in equation~\eqref{eq:2.2} 
may differ by a boundary term that could be associated with boundary contour dynamics. Here we consider periodic 
boundary conditions, $\mcl{D}$ is a two-torus, and consequently boundary terms are readily eliminated upon integrations 
by parts. Thus, $\mcl{D}$ is a rectangular domain with edges aligned along Cartesian axes and normalized to unity so 
$x,y\in[0,1)$.

The equation of motion for $\omega$ follows from the Poisson bracket according to
\begin{equation}
 \begin{aligned}
 \frac{\del\omega}{\del t}&=\{\omega,H\}
 =\mcl{J}(\omega)\frac{\delta H}{\delta\omega}\\
 &=-\left[\omega, h_1 +  \int_{\mcl{D}}\rmd^{2}r'\; h_2\, \omega\right]\,,
 \end{aligned}
 \label{eq:2.3}
\end{equation}
where the second line follows upon insertion of \eqref{dHdom}. For Euler's Hamiltonian of equation~\eqref{euler}, 
$h_{1}=0$ and $h_{2}=-\delta(r-r')\Delta^{-1}$.

Noncanonical Poisson brackets like \eqref{eq:2.2} are degenerate and this gives rise to the so-called 
Casimir invariants. An easy calculation shows that 
\begin{equation}
 \mcl{C}[\omega]=\int_{\mcl{D}} \rmd^{2}r  \, f(\omega)\,,
 \label{eq:2.4}
\end{equation}
where $f$ is an arbitrary function of $\omega$, is a constant of motion for {\it any} Hamiltonian.  
Such quantities, Casimir invariants, satisfy 
\begin{equation}
 \{F,\mcl{C}\}=0
 \label{eq:2.5}
\end{equation}
for any functional $F[\omega]$. In older plasma literature Casimir invariants were 
called generalized entropies. For convenience, the following family of Casimirs is often used:
\begin{equation}
 \mcl{C}^{(n)}[\omega]=\int_{\mcl{D}}\,\omega^{n}\,\rmd^{2}r,
 \label{eq:2.6}
\end{equation}
for $n\in\mbb{N}$. For vortex dynamics the case $n=1$ corresponds to the total vorticity, 
while $n=2$ is generally called the enstrophy. 

%%%%%%%%%%%%%%%%%%%%%%%%%%%%%%%%%%%%%%%%%%%%%%%%%%%%%%%%%%%%%%%%%%%%%%%%%%%%%%%%%%%%%%%%%%%%%%%%%%%%%%%%%%%%%%%%%%%%%%%%%%%%%%%%%%%%%%%%%%%

% Section3 %%%%%%%%%%%%%%%%%%%%%%%%%%%%%%%%%%%%%%%%%%%%%%%%%%%%%%%%%%%%%%%%%%%%%%%%%%%%%%%%%%%%%%%%%%%%%%%%%%%%%%%%%%%%%%%%%%%%%%%%%%%%%%%%

% Section3 %%%%%%%%%%%%%%%%%%%%%%%%%%%%%%%%%%%%%%%%%%%%%%%%%%%%%%%%%%%%%%%%%%%%%%%%%%%%%%%%%%%%%%%%%%%%%%%%%%%%%%%%%%%%%%%%%%%%%%%%%%%%%%%%
\section{Direct truncation}
\label{sec:3}

Equation~\eqref{eq:2.3} is an infinite-dimensional Hamiltonian system for the field~$\omega$. Our goal is to extract from it 
a finite-dimensional Hamiltonian system. We proceed by expressing \eqref{eq:2.3} in a Fourier series, which we then truncate 
to obtain a four-wave model. This is done in two parts, first for the Poisson bracket, then for the specific Hamiltonian 
of Euler's equation; however, the procedure could be carried out for any Hamiltonian. We will see that this approach leads 
to a system that is energy conserving, but it does not lead to Hamiltonian form. Our approach can be viewed as an attempt 
to obtain a Hamiltonian truncation by following the prescription of Refs.~\onlinecite{LashmoreDavies01,LashmoreDavies05} 
for Euler's equation, although in a more general setting. This section is broken up into several subsections that contain 
calculations of relevance to sections \ref{sec:4} and \ref{sec:5}, where we make comparison with a truncated system 
obtained by our beatification procedure. 

% Subsection 3.1 %%%%%%%%%%%%%%%%%%%%%%%%%%%%%%%%%%%%%%%%%%%%%%%%%%%%%%%%%%%%%%%%%%%%%%%%%%%%%%%%%%%%%%%%%%%%%%%%%%%%%%%%%%%%%%%%%%%%%%%%%%
\subsection{Reference state}
\label{ssc:3.1}

The first step of our calculation is to consider dynamics relative to  an arbitrary reference state, 
\begin{equation}
 \omega(x,y;t)=\omega_{0}(x,y)+\vep\mu(x,y;t),
 \label{eq:3.1}
\end{equation}
where $\omega_{0}$ is the reference state,  a chosen time-independent function, $\mu$ is a  new dynamical 
field, and $\vep$ is a perturbative bookkeeping parameter.  For  situations where the field $\mu$ describes 
a small deviation from  $\omega_{0}$,   which will  occur for sufficiently short time intervals, we can expand 
using $\vep\ll1$ to obtain reduced models. 

As a preparatory step for the decomposition of the quantities \eqref{euler} and \eqref{eq:2.2} into the vorticity 
Fourier amplitudes, we transform the Poisson bracket from one in terms of the  field $\omega$ to one in terms of 
$\mu$. A straightforward functional chain rule calculation (see Ref.~\onlinecite{Morrison98}) gives 
\begin{equation}
 \{F,G\}=\frac{1}{\vep^{2}}\int_{\mcl{D}}\rmd^{2}r\,
 \frac{\delta F}{\delta\mu}
 \mcl{J}_{\vep}(\mu)
 \frac{\delta G}{\delta\mu},
 \label{eq:3.3}
\end{equation}
where $\mcl{J}_{\vep}(\mu)=\mcl{J}(\omega_{0}+\vep\mu)$ is  the new Poisson operator. 

Next, inserting \eqref{eq:3.1} into the Hamiltonian of \eqref{euler} gives
\begin{equation}
 H[\mu]=-\frac{1}{2}\int_{\mcl{D}}\rmd^{2}r\,
 \left(\omega_{0}\Delta^{-1}\omega_{0}
 +2\vep\mu\Delta^{-1}\omega_{0}
 +\vep^{2}\mu\Delta^{-1}\mu\right).
 \label{eq:3.4}
\end{equation}
The Hamiltonian of \eqref{eq:3.4} together with the Poisson bracket of \eqref{eq:3.3} generates 
the exact Euler's equation. For this case we expand and project to reduce the dynamics. 

In principle, we need not select a specific form for the reference state~$\omega_{0}$ in the construction of 
most of our future results. However, when the Hamiltonian and Poisson bracket are projected onto Fourier modes, 
the following particular form of the function~$\omega_0$ is chosen for our four-wave model:
\begin{equation}
 \omega_{0}=\omega_{\alpha}\rme^{2\pi\rmi\alpha x}+\omega_{\alpha}\cg\rme^{-2\pi\rmi\alpha x}
 +\omega_{\beta}\rme^{2\pi\rmi\beta y}+\omega_{\beta}\cg\rme^{-2\pi\rmi\beta y}\,,
 \label{eq:3.2}
\end{equation}
where $\omega_{\alpha}$ and $\omega_{\beta}$ are constant complex amplitudes for modes aligned with the $x$ and 
$y$ axes, respectively. The quantities $2\pi\alpha$ and $2\pi\beta$, for $\alpha,\beta\in\mbb{Z}\setminus\!\{0\}$, 
are wave numbers for the two independent Fourier modes considered; thus, \eqref{eq:3.2} is the superposition 
of two real orthogonal waves with fixed wavelengths and zero frequency. This reference state is the simplest 
configuration that, as shown in section~\ref{sec:5}, allows the construction of a Hamiltonian model  
with four mutually interacting waves.

% Subsection 3.2 %%%%%%%%%%%%%%%%%%%%%%%%%%%%%%%%%%%%%%%%%%%%%%%%%%%%%%%%%%%%%%%%%%%%%%%%%%%%%%%%%%%%%%%%%%%%%%%%%%%%%%%%%%%%%%%%%%%%%%%%%%
\subsection{Fourier decomposition}
\label{ssc:3.2}

We perform a Fourier decomposition within the Hamiltonian description, i.e., both the Poisson bracket~\eqref{eq:3.3} 
and our example with the Hamiltonian of \eqref{euler} for vorticity dynamics are written in terms of Fourier series, 
giving a countably infinite-dimensional Hamiltonian system. We expand 
\begin{equation}
 \mu(x,y;t)=\suml_{j,k=-\infty}^{\infty}\mu_{j,k}(t)\rme^{2\pi\rmi(jx+ky)},
 \label{eq:3.5}
\end{equation}
where the amplitudes $\mu_{j,k}$ are time dependent and,  because $\mu$ is a real-valued field, satisfy 
the reality condition $\mu_{j,k}\cg=\mu_{-j,-k}$.   

Substitution of  \eqref{eq:3.5} into an arbitrary functional $F[\mu]$  and calculation of the spatial integrals
yields a function of {\it all} of the Fourier amplitudes, which we will denote by $\b{F}(\mu_{j,k})$.  Thus,  
under this  variable change  $F[\mu]=\b{F}(\mu_{j,k})$. As can be readily shown (see Ref.~\onlinecite{Morrison98}),
the derivatives of the function $\b{F}$ are related to the functional derivative of $F$ by the following identity:
\begin{equation}
 \frac{\del\b{F}}{\del \mu\cg_{j,k}}=\left(\frac{\delta F}{\delta\mu}\right)_{j,k}\,.
 \label{eq:3.6}
\end{equation}
Since $\delta F/\delta \mu$ is a function of $x$ and $y$ it can also be  Fourier expanded,   
\begin{equation}
 \frac{\delta F}{\delta\mu}=
 \suml_{j,k=-\infty}^{\infty}\left(\frac{\delta F}{\delta\mu}\right)_{j,k}
 \rme^{2\pi\rmi(jx+ky)}.
 \label{eq:3.7}
\end{equation}
Then, substitution of \eqref{eq:3.2}, \eqref{eq:3.5}, \eqref{eq:3.6}, and \eqref{eq:3.7} 
into \eqref{eq:3.3}, gives the Poisson bracket in terms of the dynamical variables $\mu_{j,k}$, 
\begin{equation}
 \begin{aligned}
 \{\b{F},\b{G}\}=&-\left(\frac{2\pi}{\vep}\right)^{2}
 \suml_{j,k=-\infty}^{\infty}\frac{\del\b{F}}{\del \mu_{j,k}\cg}
 \left[\alpha k\omega_{\alpha}\frac{\del\b{G}}{\del \mu_{(j+\alpha),k}}
 -\alpha k\omega_{\alpha}\cg\frac{\del\b{G}}{\del \mu_{(j-\alpha),k}}
 -j\beta\omega_{\beta}\frac{\del\b{G}}{\del \mu_{j,(k+\beta)}}\right.\\
 &\left.+j\beta\omega_{\beta}\cg\frac{\del\b{G}}{\del \mu_{j,(k-\beta)}}
 +\vep\suml_{m,n=-\infty}^{\infty}(km-jn)\mu_{m,n}
 \frac{\del\b{G}}{\del \mu_{(j+m),(k+n)}}\right].
 \end{aligned}
   \label{eq:3.8}
\end{equation}

Any Hamiltonian functional of the form of \eqref{eq:2.1} can be projected, $H[\mu]=\b{H}(\mu_{j,k})$, 
but for simplicity we will only consider the special case of \eqref{eq:3.4} corresponding to Euler's 
equation. Accordingly, inserting \eqref{eq:3.2} and \eqref{eq:3.5} into \eqref{eq:3.4}, yields the 
following Hamiltonian function:
\begin{equation}
 \begin{aligned}
 \b{H}=&\frac{1}{4\pi^{2}}\left[\frac{1}{\alpha^{2}}\left(
 \omega_{\alpha}\cg\omega_{\alpha}+\vep\omega_{\alpha}\mu_{\alpha,0}\cg+\vep\omega_{\alpha}\cg \mu_{\alpha,0}
 \right)+\frac{1}{\beta^{2}}\left(
 \omega_{\beta}\cg\omega_{\beta}+\vep\omega_{\beta}\mu_{0,\beta}\cg+\vep\omega_{\beta}\cg \mu_{0,\beta}\right)
 \right]\\
 &+\frac{\vep^{2}}{8\pi^{2}}\suml_{j,k=-\infty}^{\infty}\frac{\mu_{j,k}\cg \mu_{j,k}}{j^{2}+k^{2}}\,, 
 \end{aligned}
 \label{eq:3.9}
\end{equation}
where in deriving \eqref{eq:3.9} we have used the identity $f_{j,k}=-\frac{1}{(2\pi)^2}\frac{g_{j,k}}{j^2+k^2}$
which follows from $f=\Delta^{-1}g$.

Finally, with  the Hamiltonian of  \eqref{eq:3.9} and the bracket of \eqref{eq:3.8},  
the equations of motion for the Fourier amplitudes of the perturbative field are given 
in the following Hamiltonian form:
\begin{equation}
 \begin{aligned}
 \dot{\mu}_{j,k}&=\{\mu_{j,k},\b{H}\}\\
 &=\frac{1}{\vep}\suml_{m,n=-\infty}^{\infty}\frac{jn-km}{m^{2}+n^{2}}  
 \big(\omega_{m,n}+\vep \mu_{m,n}\big)\, 
 \big(\omega_{j-m,k-n}+\vep \mu_{j-m,k-n}\big),
 \end{aligned}
 \label{eq:3.10}
\end{equation}
where, in order to simplify  this expression, we introduced the definition
\begin{equation}
 \omega_{m,n}=\omega_{\alpha}\delta_{m,\alpha}\delta_{n,0}+\omega_{\alpha}\cg\delta_{m,-\alpha}\delta_{n,0}
 +\omega_{\beta}\delta_{m,0}\delta_{n,\beta}+\omega_{\beta}\cg\delta_{m,0}\delta_{n,-\beta}\,.
 \label{omegamn}
\end{equation}

% Subsection 3.3 %%%%%%%%%%%%%%%%%%%%%%%%%%%%%%%%%%%%%%%%%%%%%%%%%%%%%%%%%%%%%%%%%%%%%%%%%%%%%%%%%%%%%%%%%%%%%%%%%%%%%%%%%%%%%%%%%%%%%%%%%%
\subsection{A four-wave truncation}
\label{ssc:3.3}

So far, no approximations have been made, only a shift of the dependent variable and Fourier expansion. 
Now  we truncate \eqref{eq:3.10},  with  the objective of highlighting the major disturbances on the 
reference state for sufficiently short periods of time. For this reason, an adequate implementation 
of the truncation process must preserve the Fourier coefficients of $\mu$ representing {\it direct} 
changes of amplitudes with the same wave numbers as those of the reference state of \eqref{eq:3.2}.  
Thus, we retain the amplitudes corresponding to the following wave vectors:
\begin{subequations}
 \label{eq:3.11}
 \begin{align}
 &\vt{k}_{\alpha,0}=2\pi(\begin{array}{c c} \alpha & 0
 \end{array})\T,
 \label{eq:3.11a}\\[1.2ex]
 &\vt{k}_{0,\beta}=2\pi(\begin{array}{c c} 0 & \beta
 \end{array})\T,
 \label{eq:3.11b}\\[1.2ex]
 &\vt{k}_{-\alpha,0}=2\pi(\begin{array}{c c} -\alpha & 0
 \end{array})\T=-\vt{k}_{\alpha,0},
 \label{eq:3.11c}\\[1.2ex]
 &\vt{k}_{0,-\beta}=2\pi(\begin{array}{c c} 0 & -\beta
 \end{array})\T=-\vt{k}_{0,\beta}\,,
 \label{eq:3.11d}
 \end{align}
\end{subequations}
where $T$ denotes transpose. Notice that the Fourier amplitudes labeled by the  wave vectors $\vt{k}_{-\alpha,0}$ 
and $\vt{k}_{0,-\beta}$ are not  independent of those labeled by $\vt{k}_{\alpha,0}$ and  $\vt{k}_{0,\beta}$  
because of the reality conditions on $\omega_{0}$ and $\mu$. 

As seen from \eqref{eq:3.10}, the value of $\dot{\mu}_{j,k}$ results from a summation 
over specific quadratic terms, products of Fourier amplitudes with corresponding  wave 
vectors  summing to $\vt{k}_{j,k}=2\pi(\begin{array}{c c}j & k\end{array})\T$. Therefore, 
given that the modes associated with the wave vectors~\eqref{eq:3.11} are the only ones with 
relevant initial amplitudes in the field~$\omega$, then the only amplitudes with significant 
initial time variation are those associated with the following wave vectors\footnote{Although 
the quantities $\vt{k}_{2\alpha,0}$, $\vt{k}_{-2\alpha,0}$, $\vt{k}_{0,2\beta}$, $\vt{k}_{0,-2\beta}$ 
and $\vt{k}_{0,0}$ also represent possible sums of the vectors~\eqref{eq:3.11}, the equations of motion 
for the amplitudes $\mu_{2\alpha,0}$, $\mu_{-2\alpha,0}$, $\mu_{0,2\beta}$, $\mu_{0,-2\beta}$ and 
$\mu_{0,0}$ exhibit only identically null terms arising from the coupling between the variables 
$\mu_{\alpha,0}$, $\mu_{0,\beta}$, $\mu_{-\alpha,0}$ and $\mu_{0,-\beta}$.}:
\begin{subequations}
 \label{eq:3.12}
 \begin{align}
 &\vt{k}_{\alpha,\beta}=2\pi(\begin{array}{c c} \alpha & \beta
 \end{array})\T,
 \label{eq:3.12a}\\[1.2ex]
 &\vt{k}_{\alpha,-\beta}=2\pi(\begin{array}{c c} \alpha & -\beta
 \end{array})\T,
 \label{eq:3.12b}\\[1.2ex]
 &\vt{k}_{-\alpha,\beta}=2\pi(\begin{array}{c c} -\alpha & \beta
 \end{array})\T,
 \label{eq:3.12c}\\[1.2ex]
 &\vt{k}_{-\alpha,-\beta}=2\pi(\begin{array}{c c} -\alpha & -\beta
 \end{array})\T.
 \label{eq:3.12d}
 \end{align}
\end{subequations}
Recall, the identities $\vt{k}_{\alpha,\beta}=-\vt{k}_{-\alpha,-\beta}$ and 
$\vt{k}_{\alpha,-\beta}=-\vt{k}_{-\alpha,\beta}$  imply  $\mu_{\alpha,\beta}\cg=\mu_{-\alpha,-\beta}$ 
and $\mu_{\alpha,-\beta}\cg=\mu_{-\alpha,\beta}$; that is, the four wave vectors shown in equation~\eqref{eq:3.12} 
correspond to only two independent complex amplitudes.

Also in accordance with   \eqref{eq:3.10}, note that the amplitudes resulting from the  vectors~\eqref{eq:3.12} 
have dominant temporal variations in terms of  $\vep$.  In other words,  the differential equations for the 
velocities $\dot{\mu}_{\alpha,\beta}$ and $\dot{\mu}_{\alpha,-\beta}$ are the  only ones that have leading 
order terms independent of  the perturbative field $\mu$.  This property, together with the arguments 
mentioned in previous paragraphs, justifies the retention of  complex amplitudes associated with wave 
vectors \eqref{eq:3.11} and \eqref{eq:3.12} as dynamical  variables in  our truncation, specifically 
considering the reference state~\eqref{eq:3.2}.
 
For convenience we define  
\begin{equation}
 \begin{aligned}
 \hat{\mu}&=(\mu_{\alpha,0}, \,  \mu_{0,\beta}, \,  \mu_{\alpha,\beta}, \,  \mu_{\alpha,-\beta},\,   
 \mu_{-\alpha,0}, \,   \mu_{0,-\beta}, \,  \mu_{-\alpha,-\beta}, \,  \mu_{-\alpha,\beta})\\
 &=(\mu_{\alpha,0}, \,  \mu_{0,\beta}, \,  \mu_{\alpha,\beta}, \,  \mu_{\alpha,-\beta},\,   
 \mu_{\alpha,0}\cg, \,   \mu_{0,\beta}\cg, \,  \mu_{\alpha,\beta}\cg, \,  \mu_{\alpha,-\beta}\cg)
 \end{aligned}
 \label{eq:3.13}
\end{equation}
for the amplitudes that survive the truncation. Observe that $\hat\mu$ has eight components,  
four independent complex variables. 

Now consider the Poisson bracket. By retaining  modes with the amplitudes of \eqref{eq:3.13} 
the Poisson bracket of \eqref{eq:3.8} can be truncated to the following  bilinear operation 
between two arbitrary functions on the truncated phase space:
\begin{equation}
 \{f,g\}_{\hat{\mu}}=\left(\frac{\del f}{\del \hat{\mu}}\right)\T \!\! \cdot
 J_{\hat{\mu}}\cdot  \left(\frac{\del g}{\del \hat{\mu}}\right),
 \label{eq:3.14}
\end{equation}
\noindent where $\del/\del \hat{\mu}$ symbolizes the eight-dimensional gradient in the coordinates 
of \eqref{eq:3.13}, and the Poisson operator when truncated becomes the following matrix: 
\begin{equation}
 J_{\hat{\mu}}=-\frac{4\pi^{2}\alpha\beta}{\vep}
 \left(\begin{array}{c c c c c c c c}
 0 & \mu_{\alpha,\beta} & 0 & 0 & 0 & -\mu_{\alpha,-\beta} & -\varpi_{0,\beta}\cg & \varpi_{0,\beta} \\
 -\mu_{\alpha,\beta} & 0 & 0 & -\varpi_{\alpha,0} & \mu_{\alpha,-\beta}\cg & 0 & \varpi_{\alpha,0}\cg & 0 \\
 0 & 0 & 0 & 0 & \varpi_{0,\beta} & -\varpi_{\alpha,0} & 0 & 0 \\
 0 & \varpi_{\alpha,{0}} & 0 & 0 & -\varpi_{0,\beta}\cg & 0 & 0 & 0 \\
 0 & -\mu_{\alpha,-\beta}\cg & -\varpi_{0,\beta} & \varpi_{0,\beta}\cg & 0 & \mu_{\alpha,\beta}\cg & 0 & 0 \\
 \mu_{\alpha,-\beta} & 0 & \varpi_{\alpha,0} & 0 & -\mu_{\alpha,\beta}\cg & 0 & 0 & -\varpi_{\alpha,0}\cg \\
 \varpi_{0,\beta}\cg & -\varpi_{\alpha,0}\cg & 0 & 0 & 0 & 0 & 0 & 0 \\
 -\varpi_{0,\beta} & 0 & 0 & 0 & 0 & \varpi_{\alpha,0}\cg & 0 & 0 
 \end{array}\right),
 \label{eq:3.15}
\end{equation}
in which, for convenience, we introduced  two new auxiliary quantities 
$\varpi_{\alpha,0}=\vep^{-1}\omega_{\alpha}+\mu_{\alpha,0}$ and 
$\varpi_{0,\beta}=\vep^{-1}\omega_{\beta}+\mu_{0,\beta}$.

Next we truncate the Hamiltonian of \eqref{eq:3.9} by retaining only the eight Fourier amplitudes of $\hat{\mu}$, 
giving  
\begin{equation}
 \begin{aligned}
 \b{H}_{\hat\mu}=&\frac{1}{4\pi^{2}}\left[
 \frac{1}{\alpha^{2}}(\omega_{\alpha}\cg+\vep \mu_{\alpha,0}\cg)(\omega_{\alpha}+\vep \mu_{\alpha,0})
 +\frac{1}{\beta^{2}}(\omega_{\beta}\cg+\vep \mu_{0,\beta}\cg)(\omega_{\beta}+\vep \mu_{0,\beta})\right.\\
 &\left.+\frac{\vep^{2}}{\alpha^{2}+\beta^{2}}(\mu_{\alpha,\beta}\cg \mu_{\alpha,\beta}
 +\mu_{\alpha,-\beta}\cg \mu_{\alpha,-\beta})\right].
 \end{aligned}
 \label{eq:3.16}
\end{equation}
Similar expressions can be obtained for the Hamiltonians described in section \ref{sec:2}, 
in particular, for the Hamitlonian of \eqref{genHam} additional terms would be added to 
\eqref{eq:3.16}. 

Using the results \eqref{eq:3.14} and \eqref{eq:3.16}, the truncated equations of motion take 
the following form:
\begin{equation}
 \dot{\hat{\mu}}=J_{\hat{\mu}}\cdot \frac{\del\b{H}_{\hat{\mu}}}{\del \hat{\mu}}\,,
 \label{eq:3.17}
\end{equation}
which gives our four-wave model, 
\begin{subequations}
 \label{eq:3.18}
 \begin{align}
 &\dot{\mu}_{\alpha,0}=\alpha\beta\left(\frac{1}{\beta^{2}}-\frac{1}{\alpha^{2}+\beta^{2}}\right)
 \left[(\omega_{\beta}+\mu_{0,\beta})\mu_{\alpha,-\beta}-(\omega_{\beta}\cg+\mu_{0,\beta}\cg)\mu_{\alpha,\beta}\right],
 \label{eq:3.18a}\\[1.2ex]
 &\dot{\mu}_{0,\beta}=\alpha\beta\left(\frac{1}{\alpha^{2}}-\frac{1}{\alpha^{2}+\beta^{2}}\right)
 \left[(\omega_{\alpha}\cg+\mu_{\alpha,0}\cg)\mu_{\alpha,\beta}-(\omega_{\alpha}+\mu_{\alpha,0})\mu_{\alpha,-\beta}\cg\right],
 \label{eq:3.18b}\\[1.2ex]
 &\dot{\mu}_{\alpha,\beta}=\alpha\beta\left(\frac{1}{\beta^{2}}-\frac{1}{\alpha^{2}}\right)
 (\omega_{\alpha}+\mu_{\alpha,0})(\omega_{\beta}+\mu_{0,\beta}),
 \label{eq:3.18c}\\[1.2ex]
 &\dot{\mu}_{\alpha,-\beta}=\alpha\beta\left(\frac{1}{\alpha^{2}}-\frac{1}{\beta^{2}}\right)
 (\omega_{\alpha}+\mu_{\alpha,0})(\omega_{\beta}\cg+\mu_{0,\beta}\cg),
 \label{eq:3.18d}
 \end{align}
\end{subequations}
where we omit the equations for the complex conjugate amplitudes, since they are apparent, and we set  
$\vep=1$ because  retention of $\vep$ is not necessary, the perturbation order being the same as  
the polynomial degree of $\hat\mu$.

Alternative to the  procedure adopted above, we could have obtained the dynamical system of \eqref{eq:3.18} 
directly by truncating  equations \eqref{eq:3.10}. However, we chose to  truncate  the Poisson bracket and 
the Hamiltonian function, as these results will be important for our discussions in the following sections.

% Subsection 3.4 %%%%%%%%%%%%%%%%%%%%%%%%%%%%%%%%%%%%%%%%%%%%%%%%%%%%%%%%%%%%%%%%%%%%%%%%%%%%%%%%%%%%%%%%%%%%%%%%%%%%%%%%%%%%%%%%%%%%%%%%%%
\subsection{Constants of motion}
\label{ssc:3.4}

Having obtained our equations of motion in the form of \eqref{eq:3.17}, the question of 
which constants of motion survive the truncation arises. Because of the evident antisymmetry 
of the matrix of \eqref{eq:3.15}, it is clear that any (autonomous) Hamiltonian used to 
generate the dynamics will be conserved. Thus, this is the case for $\b{H}_{\hat{\mu}}$ of 
\eqref{eq:3.16}. However, clearly not all of the infinite number of Casimirs $\mcl{C}^{(n)}$ 
of $\eqref{eq:2.6}$ can survive, because Casimirs are associated with the null space of 
$J_{\hat{\mu}}$ which must be finite.\footnote{Equation \eqref{eq:2.5} for our truncated system
is equivalent to the condition $J_{\hat{\mu}}\cdot{\del\b{C}_{\hat{\mu}}}/{\del \hat{\mu}}\equiv 0$.} 
One can directly calculate the null eigenvectors of the matrix of \eqref{eq:3.15}, and then integrate 
a linear combination of them to obtain the following Casimir invariant:
\begin{equation}
 \begin{aligned}
 \b{\mcl{C}}^{(2)}_{\hat\mu}=&2\left[
 (\omega_{\alpha}\cg+\mu_{\alpha,0}\cg)(\omega_{\alpha}+\mu_{\alpha,0}) 
 +(\omega_{\beta}\cg+\mu_{0,\beta}\cg)(\omega_{\beta}+\mu_{0,\beta})\right.\\
 &\left.+\mu_{\alpha,\beta}\cg \mu_{\alpha,\beta}
 +\mu_{\alpha,-\beta}\cg \mu_{\alpha,-\beta}\right].
 \end{aligned}
 \label{eq:3.21}
\end{equation}
Alternatively, one expects the quadratic Casimir to survive, it being a so-called rugged invariant.\cite{Montgomery79}   
Thus, inserting the transformation \eqref{eq:3.1} into \eqref{eq:2.6} for $n=2$ gives the candidate 
\begin{equation}
 \mcl{C}^{(2)}[\mu]=\int_{\mcl{D}}\rmd^{2}r \left(\omega_{0}^{2}+2\vep\omega_{0}\mu+\vep^{2}\mu^2\right).
 \label{eq:3.19}
\end{equation}
 Employing the expansion \eqref{eq:3.5} to the above yields
\begin{equation}
 \b{\mcl{C}}^{(2)}=2(\omega_{\alpha}\cg\omega_{\alpha}+\omega_{\beta}\cg\omega_{\beta}
 +\omega_{\alpha}\mu_{\alpha,0}\cg+\omega_{\alpha}\cg \mu_{\alpha,0}
 +\omega_{\beta}\mu_{0,\beta}\cg+\omega_{\beta}\cg \mu_{0,\beta})
 +\suml_{j,k=-\infty}^{\infty}\mu_{j,k}\cg \mu_{j,k},
 \label{eq:3.20}
\end{equation}
where the parameter~$\vep$ was set to unit. Then upon truncating \eqref{eq:3.20}, 
i.e., retaining only the complex amplitudes present in \eqref{eq:3.13}, indeed 
we obtain \eqref{eq:3.21}.

% Subsection 3.5 %%%%%%%%%%%%%%%%%%%%%%%%%%%%%%%%%%%%%%%%%%%%%%%%%%%%%%%%%%%%%%%%%%%%%%%%%%%%%%%%%%%%%%%%%%%%%%%%%%%%%%%%%%%%%%%%%%%%%%%%%%
\subsection{The Jacobi identity}
\label{ssc:3.5}

So far, we have performed a truncation of the Hamiltonian formulation for the two-dimensional Euler equation, 
yielding the four-wave dynamical system of equations  \eqref{eq:3.18a}--\eqref{eq:3.18d}. En route we obtained 
the invariant function $\b{H}_{\hat\mu}$ and the bilinear operation of \eqref{eq:3.14}. However, the resulting 
system, although energy conserving, cannot be said to be Hamiltonian unless  the $J_{\hat{\mu}}$ of \eqref{eq:3.15} 
when inserted into \eqref{eq:3.14} produces a bracket that satisfies the Jacobi identity. In the present section 
we briefly review features of finite-dimensional noncanonical Hamiltonian systems, present the Jacobi identity, 
and discuss its failure for $J_{\hat{\mu}}$. 

In conventional physics texts, Hamiltonian dynamics is presented in terms of canonical coordinates and momenta, 
for which a coordinate free geometric approach\cite{Jost64,Abraham87} is available. Alternatively, one can 
consider a noncanonical Hamiltonian framework based on the Lie algebraic properties of the Poisson bracket  
(e.g., Refs.~\onlinecite{Sudarshan74, Morrison82b,Morrison98}), where coordinates need not be canonical 
and degeneracy in the Poisson bracket is allowed. (See,  e.g., Ref.\ \onlinecite{Weinstein83} for geometrical 
description.) Such a formulation occurs naturally  in a variety of contexts, notably the Eulerian variable 
description of matter, where many fluid and plasma applications have been treated,\cite{Chandre14,Morrison98,
Morrison80,Morrison82a,Morrison82b} and also in the context of semiclassical approximations with generalized 
coherent states.\cite{Viscondi11a,Viscondi11b}

We consider a space (manifold) with $N$ real\footnote{Alternatively, we could employ complex
coordinates and their conjugate values, in a similar way to the results of subsections \ref{ssc:3.2} 
and \ref{ssc:3.3}.} coordinates, $z=(z^{1},z^{2},\ldots,z^{N})$, and  define a bilinear operation 
between two arbitrary functions as follows: 
\begin{equation}
 \{f,g\}=\left(\frac{\del f}{\del z}\right)\T\!\! \!\cdot J(z)\cdot \frac{\del g}{\del z},
 \label{eq:4.2}
\end{equation}
where $\del/\del z$ represents the gradient in the coordinates $z$ and $J(z)$ in \eqref{eq:4.2} 
is a matrix with possible functional dependence on  $z$.

Thus far no restrictions have been placed on the matrix $J(z)$; however, in order for \eqref{eq:4.2}
to be a Poisson bracket, two additional conditions are required. First, it must be antisymmetric
\begin{equation}
 \{f,g\}=-\{g,f\}\,, 
 \label{eq:4.3}
\end{equation}
and second it  must satisfy the Jacobi identity,
\begin{equation}
 \{f,\{g,h\}\}+\{g,\{h,f\}\}+\{h,\{f,g\}\}=0\,.
 \label{eq:4.4}
\end{equation}
Properties \eqref{eq:4.3} and \eqref{eq:4.4} imply conditions on the matrix $J(z)$, viz. 
\begin{subequations}
 \label{eq:4.5}
 \begin{align}
 &J^{ab}=-J^{ba}
 \label{eq:4.5a}\\[1.2ex]
 &J^{ad}\frac{\del J^{bc}}{\del z^{d}}
 +J^{bd}\frac{\del J^{ca}}{\del z^{d}}
 +J^{cd}\frac{\del J^{ab}}{\del z^{d}}=0,
 \label{eq:4.5b}
 \end{align}
\end{subequations}
which if true for all  $a,b,c=1,2,\ldots,N$ are equivalent to \eqref{eq:4.3} and \eqref{eq:4.4}.  
Note, in \eqref{eq:4.5b} repeated sum notation is assumed. When both of  the above conditions 
are met, we call $J(z)$ a Poisson matrix. Note that \eqref{eq:4.5b} is immediately satisfied 
by a matrix with no dependence on the variables $z$. That is, a skew-symmetric matrix with 
constant elements automatically produces a Poisson bracket.

Given a Poisson matrix $J(z)$, the Hamiltonian equations of motion are 
\begin{equation}
 \dot{z}^{a}=\{z^{a}, {H}\}=  J^{ab}\frac{\del {H}}{\del z^{b}},
 \label{eq:4.6}
\end{equation}
for $a,b=1,2,\ldots,N$, where  the function ${H}(z)$  is the  {Hamiltonian}.

The definition~\eqref{eq:4.6} is a quite general (coordinate dependent) formulation of a  
Hamiltonian system. In particular, we  say that the dynamical system~\eqref{eq:4.6} is 
in canonical form if its Poisson matrix is 
\begin{equation}
 J_{c}=\left(\begin{array}{c c}
 0_{r\times r} & \mds{1}_{r\times r} \\
 -\mds{1}_{r\times r} & 0_{r\times r}
 \end{array}\right),
 \label{eq:4.7}
\end{equation}
where  $0_{r\times r}$ and $\mds{1}_{r\times r}$ denote, respectively, the zero 
and identity matrices of order $r=N/2$, canonical systems being even dimensional.

Returning to the case at hand, the matrix of \eqref{eq:3.15}, we have demonstrated 
by inserting this $J_{\hat\mu}$ into the left-hand-side of \eqref{eq:4.5b} its failure 
to vanish. Therefore, \eqref{eq:3.14} is not a Poisson bracket, and  the equations of 
motion \eqref{eq:3.18} are not a Hamiltonian system with \eqref{eq:3.16} as Hamiltonian.  
This failure of the Jacobi identity is not surprising, since it has been known for some  
time that direct Fourier truncation destroys the Jacobi identity.\cite{Morrison81a}

There is a caveat to our result. Strictly speaking, we have not demonstrated the absence 
of any Hamiltonian formulation for the system \eqref{eq:3.18} -- we have only shown 
that the elements of identity~\eqref{eq:3.17} do not define a Hamiltonian system. We 
cannot exclude the possibility that equations \eqref{eq:3.18} could result from some  
unknown Poisson matrix together with some invariant Hamiltonian function. However, 
since the system~\eqref{eq:3.18} follows from the truncation of a Hamiltonian model, 
we believe the existence of a Hamiltonian formulation arising from quantities 
uncorrelated with the truncated values of the expressions \eqref{eq:3.3} and  
\eqref{eq:3.4} is unlikely. Moreover, we have not been able to find any 
additional invariants that might serve as candidate Hamiltonians.   

Another feature of the failure of the Jacobi identity is worth mentioning. Because \eqref{eq:3.15} 
is antisymmetric, its rank must be even, which in our case is six. For Hamiltonian systems, the 
existence of two null eigenvectors implies the existence of two Casimir invariants; a consequence 
of the Jacobi identity is that the null space of the Poisson matrix is spanned by gradients of 
Casimir invariants. However, for the $J_{\hat\mu}$ of \eqref{eq:3.15}, there is only one independent 
function whose gradient is a null eigenvector, even though \eqref{eq:3.15} has a two-dimensional 
null space. This is another manifestation of the fact that the matrix~$J_{\hat\mu}$ does not 
satisfy the Jacobi identity. 

%%%%%%%%%%%%%%%%%%%%%%%%%%%%%%%%%%%%%%%%%%%%%%%%%%%%%%%%%%%%%%%%%%%%%%%%%%%%%%%%%%%%%%%%%%%%%%%%%%%%%%%%%%%%%%%%%%%%%%%%%%%%%%%%%%%%%%%%%%%

% Section4 %%%%%%%%%%%%%%%%%%%%%%%%%%%%%%%%%%%%%%%%%%%%%%%%%%%%%%%%%%%%%%%%%%%%%%%%%%%%%%%%%%%%%%%%%%%%%%%%%%%%%%%%%%%%%%%%%%%%%%%%%%%%%%%%

% Section4 %%%%%%%%%%%%%%%%%%%%%%%%%%%%%%%%%%%%%%%%%%%%%%%%%%%%%%%%%%%%%%%%%%%%%%%%%%%%%%%%%%%%%%%%%%%%%%%%%%%%%%%%%%%%%%%%%%%%%%%%%%%%%%%%
\section{Beatification}
\label{sec:4}

Now we perform the beatification procedure,\cite{MorrisonVanneste} a perturbative transformation  
that removes the functional dependence of the Poisson operator on the field variable and replaces 
it with a reference state. The procedure is applied to the bracket of \eqref{eq:3.3} and, 
in preparation for the truncation procedure of section \ref{sec:5}, the Hamiltonian for 
the two-dimensional Euler equation is expressed in terms of the transformed variable.   

The beatification procedure has two parts. The first part involves the Poisson bracket, with the original 
field shifted by introducing a sum of a reference state and a perturbative field as was done in equation~\eqref{eq:3.1},   
followed by an additional transformation of the Poisson bracket for the purpose of removing the field dependence  
in the Poisson operator to within a predetermined order of perturbation. The second part is to apply the 
same transformations to the Hamiltonian of interest, which in our case will be that for Euler's equation.

In the original formulation of the beatification procedure,\cite{MorrisonVanneste} the reference function was chosen  
to be an equilibrium state.   For example, for   Euler's equation we could choose  a reference state consisting of 
a single Fourier spatial mode, in contrast to the choice made in section~\ref{sec:3}. However, the exclusion of a 
spatial mode from identity~\eqref{eq:3.2} would result in restricting the dynamics of the beatified perturbative  
field to the Fourier subspace orthogonal to the reference function. That is, there would be no temporal variation 
in the perturbative  coefficient corresponding to the same wave vector of the single-mode reference state. For this 
reason, it is necessary to   modify the standard prescription for beatification, in order to obtain our  beatified 
four-wave model with similar characteristics  to the system~\eqref{eq:3.18}.\footnote{The removal of restrictions 
on the choice of the reference state also removes several simplifications of the intermediate calculations for 
obtaining the beatified equations of motion. As mentioned above equation~\eqref{eq:4.8}, if the reference state 
were an equilibrium we would obtain a dynamical system that is accurate to one higher perturbative order.}  

A penalty paid for the reference state not being an equilibrium is that the beatification needs 
to be carried out to one higher order to retain consistent nonlinearity. Thus, we introduce the 
following near-identity second-order transformation:
\begin{equation}
 \eta=\mu+\frac{\vep}{2}D\mu^{2}+\frac{\vep^{2}}{6}D^{2}\mu^{3},
 \label{eq:4.8}
\end{equation}
in which the new variable  $\eta=\eta(x,y,t)$ stands for the beatified perturbative field 
and the operator $D$ is defined by
\begin{equation}
 Df=-\frac{1}{2}\left(\frac{\del}{\del x}\frac{f}{\omega_{x}}
 +\frac{\del}{\del y}\frac{f}{\omega_{y}}\right),
 \label{eq:4.9}
\end{equation}
where $f$ is an arbitrary function, $\omega_{x}=\del_{x}\omega_{0}$, and $\omega_{y}=\del_{y}\omega_{0}$.

As a preliminary step before effecting the transformation~\eqref{eq:4.8} of the Poisson bracket \eqref{eq:3.3}, 
we write the inverse relation between the perturbative fields up to second order in the parameter~$\vep$:
\begin{equation}
 \mu=\eta-\frac{\vep}{2}D\eta^{2}
 +\frac{\vep^{2}}{2}D\eta D\eta^{2}
 -\frac{\vep^{2}}{6}D^{2}\eta^{3}+O(\vep^{3})\,.
 \label{eq:4.10}
\end{equation}

To transform the Poisson bracket we introduce the functional transformation $F[\mu]=\til{F}[\eta]$, 
which upon variation gives 
\begin{equation}
 \delta F[\mu;\delta\mu]=\int_{\mcl{D}} \rmd^{2}r\, \frac{\delta F}{\delta\mu}\delta\mu
 =\delta \til{F}[\eta;\delta\eta]=\int_{\mcl{D}} \rmd^{2}r\, \frac{\delta\til{F}}{\delta\eta}\delta\eta\,.
 \label{chainrule}
\end{equation}
Then upon varying \eqref{eq:4.8} and inserting $\delta\eta$ into the above, 
followed by integrations by parts, gives
\begin{equation}
 \frac{\delta F}{\delta\mu}=\left[1+\vep\mu D\dg
 +\frac{\vep^{2}}{2}\mu^{2}(D\dg)^{2}
 \right]\frac{\delta\til{F}}{\delta\eta}
 =:\mathcal{S}\frac{\delta\til{F}}{\delta\eta}\,. 
 \label{eq:4.11}
\end{equation}
where $D\dg$ denotes the adjoint operator of $D$ with respect to the scalar product
\begin{equation}
 \bra f,g\ket=\int_{\mcl{D}} \rmd^{2}r\, fg\,,
 \label{eq:4.12}
\end{equation}
which is defined for  two arbitrary functions on the domain $\mcl{D}$. For future reference, 
we note the action of the operator $D\dg$ on a function $f$ is given by the following formula:
\begin{equation}
 D\dg f=\frac{1}{2}\left(\frac{1}{\omega_{x}}\frac{\del f}{\del x}
 +\frac{1}{\omega_{y}}\frac{\del f}{\del y}\right).
 \label{eq:4.13}
\end{equation}
The calculation leading to \eqref{eq:4.11} amounts to  the   chain rule for functionals, and we 
refer the reader to Ref.~\onlinecite{Morrison98} for more details. 

Substitution of \eqref{eq:4.11} and the counterpart for $G$ into  
\eqref{eq:3.3} gives the transformed bracket
\begin{equation}
\{F,G\}= \frac{1}{\vep^{2}}\int_{\mcl{D}}\rmd^{2}r\,\frac{\delta F}{\delta\eta}\, 
 \mathcal{S}^{\dg}\mcl{J}_{\vep}(\mu)\mathcal{S}\, \frac{\delta G}{\delta\eta}\,,
 \label{eq:transBkt}
\end{equation}
where we have dropped the tildes on the functionals. In appendix~\ref{app:beat} 
we show that the beatified Poisson bracket is given by 
\begin{equation}
 \{F,G\}=\frac{1}{\vep^{2}}\int_{\mcl{D}}\rmd^{2}r\,\frac{\delta F}{\delta\eta}
 \mcl{J}(\omega_{0})\frac{\delta G}{\delta\eta}
 +O\left(\vep\frac{\delta F}{\delta\eta}\frac{\delta G}{\delta\eta}\right); 
 \label{eq:4.14}
\end{equation}
that is, the transformed Poisson operator $\mathcal{S}^{\dg}\mcl{J}_{\vep}(\mu)\mathcal{S}$ 
is flattened to second order by the transformation \eqref{eq:4.8}.

As expected, after the beatification procedure, the Poisson bracket consists of a Poisson operator 
that is independent of the field variable, since $\mcl{J}(\omega_{0})f=-\left[\omega_{0},f\right]$ 
for any function~$f$. Note that the applicability of the beatified Poisson bracket is limited by 
the leading order of the quantities $\delta F/\delta\eta$ and $\delta G/\delta\eta$ with respect 
to perturbative parameter~$\vep$,\footnote{That is, since the functions $\delta F/\delta\eta$ and 
$\delta G/\delta\eta$ may depend on the parameter~$\vep$ (e.g. equation~\eqref{eq:4.17}), the leading 
orders of these functional derivatives must be taken into account when determining the perturbative 
orders in which the first term on the right-hand side of identity~\eqref{eq:4.14} is valid.}
as indicated by the second term on the right-hand side of equation~\eqref{eq:4.14}.

In a manner similar to the Poisson bracket, we can also rewrite the Hamiltonian functional in terms 
of the beatified field. Substituting the transformation~\eqref{eq:4.10} into the equation~\eqref{eq:3.4}, 
we obtain the following result:
\begin{equation}
 \begin{aligned}
 H[\eta]=&-\frac{1}{2}\int_{\mcl{D}}\rmd^{2}r\,\left\{
 \omega_{0}\Delta^{-1}\omega_{0}+2\vep(\Delta^{-1}\omega_{0})\eta
 -\vep^{2}(D\dg\Delta^{-1}\omega_{0})\eta^{2}+\vep^{2}\eta\Delta^{-1}\eta\right.\\
 &\left.+\vep^{3}(D\dg\Delta^{-1}\omega_{0})(D\eta^{2})\eta
 -\frac{\vep^{3}}{3}[(D\dg)^{2}\Delta^{-1}\omega_{0}]\eta^{3}
 -\vep^{3}\eta\Delta^{-1}D\eta^{2}\right\}+O(\vep^{4}).
 \end{aligned}
 \label{eq:4.15}
\end{equation}

Given the beatified expressions for the Hamiltonian and the Poisson operator, we can readily 
write the equation of motion for the field~$\eta$:
\begin{equation}
 \frac{\del\eta}{\del t}=\{\eta,H\}
 =\frac{1}{\vep^{2}}\mcl{J}(\omega_{0})\frac{\delta H}{\delta\eta}+O(\vep^{2}),
 \label{eq:4.16}
\end{equation}
in which the functional derivative of the Hamiltonian of \eqref{eq:4.15} is given by 
\begin{equation}
 \begin{aligned}
 \frac{\delta H}{\delta\eta}=&-\vep\Delta^{-1}\omega_{0}
 +\vep^{2}\eta D\dg\Delta^{-1}\omega_{0}-\vep^{2}\Delta^{-1}\eta
 -\frac{\vep^{3}}{2}(D\dg\Delta^{-1}\omega_{0})(D\eta^{2})
 +\frac{\vep^{3}}{2}\Delta^{-1}D\eta^{2}\\
 &-\vep^{3}\eta D\dg\eta D\dg\Delta^{-1}\omega_{0}
 +\frac{\vep^{3}}{2}\eta^{2}(D\dg)^{2}\Delta^{-1}\omega_{0}
 +\vep^{3}\eta D\dg\Delta^{-1}\eta+O(\vep^{4}).
 \end{aligned}
 \label{eq:4.17}
\end{equation}

As shown in section~\ref{sec:2}, the two-dimensional Euler equation is a nonlinear dynamical system, 
due to its quadratic dependence on the field variable. Furthermore, during the presentation of the 
Hamiltonian formulation for the equation of motion~\eqref{eq:2.3}, we showed that the integrand of 
the Hamiltonian functional~\eqref{euler} corresponds to a quadratic function of the scalar vorticity, 
while the Poisson operator $\mcl{J}(\omega)$ displays linear dependence on the field, and together 
they produce the quadratic nonlinearity. Analogous to the original two-dimensional Euler equation, 
the beatified dynamical system is also a system with quadratic nonlinearity, as indicated by the 
identities~\eqref{eq:4.16} and \eqref{eq:4.17}. However, unlike the nonperturbative formulation, 
the integrand of the Hamiltonian functional~\eqref{eq:4.15} is a cubic function of the beatified 
field, while the Poisson operator $\mcl{J}(\omega_{0})$ is independent of the dynamical variable. 
Therefore, the beatification procedure transfers nonlinearity from the Poisson bracket to the 
Hamiltonian. 

%%%%%%%%%%%%%%%%%%%%%%%%%%%%%%%%%%%%%%%%%%%%%%%%%%%%%%%%%%%%%%%%%%%%%%%%%%%%%%%%%%%%%%%%%%%%%%%%%%%%%%%%%%%%%%%%%%%%%%%%%%%%%%%%%%%%%%%%%%%

% Section5 %%%%%%%%%%%%%%%%%%%%%%%%%%%%%%%%%%%%%%%%%%%%%%%%%%%%%%%%%%%%%%%%%%%%%%%%%%%%%%%%%%%%%%%%%%%%%%%%%%%%%%%%%%%%%%%%%%%%%%%%%%%%%%%%

% Section5 %%%%%%%%%%%%%%%%%%%%%%%%%%%%%%%%%%%%%%%%%%%%%%%%%%%%%%%%%%%%%%%%%%%%%%%%%%%%%%%%%%%%%%%%%%%%%%%%%%%%%%%%%%%%%%%%%%%%%%%%%%%%%%%%
\section{Truncation of the beatified system}
\label{sec:5}

Having obtained the beatified system of section \ref{sec:4}, we are set to follow the procedures 
of subsections~\ref{ssc:3.2} and \ref{ssc:3.3} to  perform  modal  decomposition followed by  
four-wave truncation, which we do in subsections~\ref{ssc:5.1} and \ref{ssc:5.2}. However in 
this case, because the starting Hamiltonian system is beatified,  the resulting truncated system 
is a Hamiltonian system. In section \ref{ssc:5.3} we present a brief discussion on the constants 
of motion for the truncated system, where it is seen, contrary to the system of section \ref{sec:3}, 
that  there are two Casimir invariants. In a companion appendix~\ref{app:canV}, we canonize the 
four-wave model by presenting the explicit transformation to  canonical  variables for the four-wave 
system. To remove clutter from some rather cumbersome equations, in subsections~\ref{ssc:5.1} and 
\ref{ssc:5.2}, we have  set our bookkeeping parameter $\vep=1$.  
 
% Subsection 5.1 %%%%%%%%%%%%%%%%%%%%%%%%%%%%%%%%%%%%%%%%%%%%%%%%%%%%%%%%%%%%%%%%%%%%%%%%%%%%%%%%%%%%%%%%%%%%%%%%%%%%%%%%%%%%%%%%%%%%%%%%%%
\subsection{Fourier decomposition}
\label{ssc:5.1}

Beginning as in subsection \ref{ssc:3.2}, we  Fourier expand $\eta$, the beatified perturbative 
field, as in \eqref{eq:3.5} with complex amplitudes $\eta_{j,k}$ satisfying the reality conditions 
$\eta_{-j,-k}=\eta_{j,k}\cg$. Then, as in subsection~\ref{ssc:3.2}, we insert the expansion into 
functionals giving $F[\eta]=\b{F}(\eta_{j,k})$, and  analogous to \eqref{eq:3.6} we have  
\begin{equation}
 \frac{\del\b{F}}{\del \eta\cg_{j,k}}=\left(\frac{\delta F}{\delta\eta}\right)_{j,k}\,.
 \label{eq:5.3}
\end{equation}
Upon substituting expressions~\eqref{eq:3.2} and \eqref{eq:5.3} into \eqref{eq:4.14}, we 
obtain the following expression for the leading order beatified Poisson bracket in terms 
of variables $\eta_{j,k}$:
\begin{equation}
 \begin{aligned}
 \{\b{F},\b{G}\}=&-(2\pi)^{2}\suml_{j,k=-\infty}^{\infty}
 \left\{\alpha k\frac{\del\b{F}}{\del \eta_{j,k}\cg}
 \left[\omega_{\alpha}\frac{\del\b{G}}{\del \eta_{(j+\alpha),k}}
 -\omega_{\alpha}\cg\frac{\del\b{G}}{\del \eta_{(j-\alpha),k}}\right]
 \right.\\
 &\left.-j\beta\frac{\del\b{F}}{\del \eta_{j,k}\cg}\left[
 \omega_{\beta}\frac{\del\b{G}}{\del \eta_{j,(k+\beta)}}
 -\omega_{\beta}\cg\frac{\del\b{G}}{\del \eta_{j,(k-\beta)}}
 \right]\right\}\,. 
 \end{aligned}
 \label{eq:5.4}
\end{equation}
 
Next we rewrite the Hamiltonian functional in terms of the $\eta_{j,k}$ by substituting \eqref{eq:3.2} 
and the Fourier expansion for $\eta$ into  \eqref{eq:4.15}. This yields the following complicated 
expression for our beatified Euler Hamiltonian to the desired perturbative order:
\begin{equation}
 \begin{aligned}
 \b{H}=&\;\frac{1}{(2\pi)^{2}}
 \left[\frac{1}{\alpha^{2}}\left(\omega_{\alpha}\eta_{\alpha,0}\cg
 +\omega_{\alpha}\cg \eta_{\alpha,0}+\omega_{\alpha}\cg\omega_{\alpha}\right)
 +\frac{1}{\beta^{2}}\left(\omega_{\beta}\eta_{0,\beta}\cg
 +\omega_{\beta}\cg \eta_{0,\beta}+\omega_{\beta}\cg\omega_{\beta}\right)\right]\\
 &+\frac{1}{(4\pi)^{2}}\suml_{j,k=-\infty}^{\infty}\left(
 \frac{2}{j^{2}+k^{2}}-\frac{1}{\alpha^{2}}-\frac{1}{\beta^{2}}
 \right)\eta_{j,k}\cg \eta_{j,k}\\
 &+\frac{1}{(8\pi)^{2}}\suml_{j,k=-\infty}^{\infty}\,
 \suml_{m,n=-\infty}^{\infty}\,\suml_{r=-\infty}^{\infty}
 \theta_{r}\eta_{j,k}\eta_{m,n}\\
 &\times\left[\kappa_{\alpha}^{j,k,m,n}
 \frac{(\omega_{\alpha}\cg)^{r}}{\omega_{\alpha}^{r+1}}
 \eta_{j+m-(2r+1)\alpha,k+n}\cg
 +\kappa_{\beta}^{j,k,m,n}
 \frac{(\omega_{\beta}\cg)^{r}}{\omega_{\beta}^{r+1}}
 \eta_{j+m,k+n-(2r+1)\beta}\cg\right]\,.
 \end{aligned}
 \label{eq:5.5}
\end{equation}
In equation \eqref{eq:5.5}, we have introduced the following definitions for constants:
\begin{equation}
 \theta_{r}=\left\{
 \begin{aligned}
 &+1,\;\mbox{if}\; r\geq0,\\
 &-1,\;\mbox{if}\; r<0,
 \end{aligned}
 \right.
 \label{eq:5.6}
\end{equation}
\begin{subequations}
 \label{eq:5.7}
 \begin{align}
 &\kappa_{\alpha}^{j,k,m,n}=
 \left\{\frac{2}{\left[j+m-(2r+1)\alpha\right]^{2}+(k+n)^{2}}
 -\frac{1}{\alpha^{2}}-\frac{1}{\beta^{2}}\right\}
 \frac{j+m-(2r+1)\alpha}{\alpha},
 \label{eq:5.7a}\\[1.2ex]
 &\kappa_{\beta}^{j,k,m,n}=
 \left\{\frac{2}{(j+m)^{2}+\left[k+n-(2r+1)\beta\right]^{2}}
 -\frac{1}{\alpha^{2}}-\frac{1}{\beta^{2}}\right\}
 \frac{k+n-(2r+1)\beta}{\beta}.
 \label{eq:5.7b}
 \end{align}
\end{subequations}

Finally, using \eqref{eq:5.4} and \eqref{eq:5.5} the beatified system assumes the Hamiltonian form, 
\begin{equation}
 \dot{\eta}_{j,k}=\{\eta_{j,k},\b{H}\}.
 \label{eq:5.8}
\end{equation}
Equation~\eqref{eq:5.8} could be written out explicitly, but we refrain from doing 
so because it is bulky and not necessary for our future development. Note however, 
as anticipated in section~\ref{sec:4}, the bracket $\{\eta_{j,k},\b{H}\}$ is  
quadratic in  $\eta_{j,k}$, since the beatified Poisson operator is independent 
of $\eta_{j,k}$ and the Hamiltonian $\b{H}(\eta_{j,k})$ is cubic in $\eta_{j,k}$.  
Thus, by effecting the beatification transformation to second order, we obtained 
a consistent system that includes all terms of quadratic order. We note in passing, 
if  our reference state  had been  an equilibrium, then our transformation could 
yield equations of motion correct to cubic order, but we will not pursue this here. 
 
% Subsection 5.2 %%%%%%%%%%%%%%%%%%%%%%%%%%%%%%%%%%%%%%%%%%%%%%%%%%%%%%%%%%%%%%%%%%%%%%%%%%%%%%%%%%%%%%%%%%%%%%%%%%%%%%%%%%%%%%%%%%%%%%%%%%
\subsection{Truncation}
\label{ssc:5.2}

In order to truncate the beatified system of \eqref{eq:5.8} for the two-dimensional Euler equation, 
we follow the procedure of subsection~\ref{ssc:3.3}. Thus, we retain the amplitudes $\eta_{j,k}$ 
that are labeled by the wave vectors of \eqref{eq:3.11} and \eqref{eq:3.12}. In analogy to 
equation \eqref{eq:3.13}, we introduce the following variable for the beatified complex amplitudes:
\begin{equation}
 \begin{aligned}
 \hat\eta&=( \eta_{\alpha,0},\,  \eta_{0,\beta},\,   \eta_{\alpha,\beta},\,    \eta_{\alpha,-\beta},\,  
 \eta_{-\alpha,0},\,   \eta_{0,-\beta},\,   \eta_{-\alpha,-\beta},\,   \eta_{-\alpha,\beta})\\
 &=( \eta_{\alpha,0},\,  \eta_{0,\beta},\,   \eta_{\alpha,\beta},\,    \eta_{\alpha,-\beta},\,  
 \eta_{\alpha,0}\cg,\,   \eta_{0,\beta}\cg,\,   \eta_{\alpha,\beta}\cg,\,   \eta_{\alpha,-\beta}\cg) \,, 
 \end{aligned}
 \label{eq:5.9}
\end{equation}
which consists of four independent complex or eight real variables. With the choice of amplitudes 
of \eqref{eq:5.9}, we proceed to the truncation of the beatified Hamiltonian system. 

First, by restricting to the variables of \eqref{eq:5.9}, the beatified Poisson bracket reduces to  
\begin{equation}
 \{f,g\}_{\hat\eta}=\left(\frac{\del f}{\del \hat\eta}\right)\T\!\!\!  \cdot 
 J_{\hat\eta} \cdot \left(\frac{\del g}{\del \hat\eta}\right),
 \label{eq:5.10}
\end{equation}
with the matrix
\begin{equation}
 J_{\hat\eta}=-(2\pi)^{2}\alpha\beta
 \left(\begin{array}{c c c c c c c c}
 0 & 0 & 0 & 0 & 0 & 0 & -\omega_{\beta}\cg & \omega_{\beta} \\
 0 & 0 & 0 & -\omega_{\alpha} & 0 & 0 & \omega_{\alpha}\cg & 0 \\
 0 & 0 & 0 & 0 & \omega_{\beta} & -\omega_{\alpha} & 0 & 0 \\
 0 & \omega_{\alpha} & 0 & 0 & -\omega_{\beta}\cg & 0 & 0 & 0 \\
 0 & 0 & -\omega_{\beta} & \omega_{\beta}\cg & 0 & 0 & 0 & 0 \\
 0 & 0 & \omega_{\alpha} & 0 & 0 & 0 & 0 & -\omega_{\alpha}\cg \\
 \omega_{\beta}\cg & -\omega_{\alpha}\cg & 0 & 0 & 0 & 0 & 0 & 0 \\
 -\omega_{\beta} & 0 & 0 & 0 & 0 & \omega_{\alpha}\cg & 0 & 0 
 \end{array}\right).
 \label{eq:5.11}
\end{equation}
Because $J_{\hat\eta}$ is antisymmetric and does not depend on $\hat\eta$, it follows 
from subsection~\ref{ssc:3.5} that it satisfies the Jacobi identity. Unlike the matrix  
of \eqref{eq:3.15} obtained by direct truncation, any antisymmetric reduction of the 
beatified bilinear operation~\eqref{eq:5.4} results in a Poisson bracket. 

It remains to obtain the Hamiltonian for the reduced system by truncation of \eqref{eq:5.5}.  
This is done by restricting $\b{H}$ of \eqref{eq:5.5} to $\hat\eta$, yielding 
\begin{equation}
 \begin{aligned}
 \b{H}_{\hat\eta}=&\frac{1}{4\pi^{2}}
 \left[\frac{1}{\alpha^{2}}\left(
 \omega_{\alpha}\eta_{\alpha,0}\cg
 +\omega_{\alpha}\cg \eta_{\alpha,0}
 +\omega_{\alpha}\cg\omega_{\alpha}\right)
 +\frac{1}{\beta^{2}}\left(
 \omega_{\beta}\eta_{0,\beta}\cg
 +\omega_{\beta}\cg \eta_{0,\beta}
 +\omega_{\beta}\cg\omega_{\beta}\right)\right]\\
 &+\frac{1}{8\pi^{2}}
 \left[\chi_{\alpha,\beta}\left(\eta_{\alpha,0}\cg \eta_{\alpha,0}-\eta_{0,\beta}\cg \eta_{0,\beta}\right)
 +\xi_{\alpha,\beta}\left(\eta_{\alpha,\beta}\cg \eta_{\alpha,\beta}+\eta_{\alpha,-\beta}\cg \eta_{\alpha,-\beta}\right)\right]\\
 &-\frac{\xi_{\alpha,\beta}}{16\pi^{2}}\left[\frac{\eta_{\alpha,\beta}\eta_{\alpha,-\beta}}{\omega_{\alpha}^{2}}
 \left(\omega_{\alpha}\eta_{\alpha,0}\cg+\omega_{\alpha}\cg \eta_{\alpha,0}\right)
 +\frac{\eta_{\alpha,\beta}\cg \eta_{\alpha,-\beta}\cg}{(\omega_{\alpha}\cg)^{2}}
 \left(\omega_{\alpha}\cg \eta_{\alpha,0}+\omega_{\alpha}\eta_{\alpha,0}\cg\right)\right.\\
 &\left.+\frac{\eta_{\alpha,\beta}\eta_{\alpha,-\beta}\cg}{\omega_{\beta}^{2}}
 \left(\omega_{\beta}\eta_{0,\beta}\cg+\omega_{\beta}\cg \eta_{0,\beta}\right)
 +\frac{\eta_{\alpha,\beta}\cg \eta_{\alpha,-\beta}}{(\omega_{\beta}\cg)^{2}}
 \left(\omega_{\beta}\cg \eta_{0,\beta}+\omega_{\beta}\eta_{0,\beta}\right)\right]\\
 &+\frac{\chi_{\alpha,\beta}}{32\pi^{2}}\left[\frac{\eta_{\alpha,\beta}\eta_{\alpha,-\beta}}{\omega_{\alpha}^{2}}
 \left(\omega_{\alpha}\eta_{\alpha,0}\cg-\omega_{\alpha}\cg \eta_{\alpha,0}\right)
 +\frac{\eta_{\alpha,\beta}\cg \eta_{\alpha,-\beta}\cg}{(\omega_{\alpha}\cg)^{2}}
 \left(\omega_{\alpha}\cg \eta_{\alpha,0}-\omega_{\alpha}\eta_{\alpha,0}\cg\right)\right.\\
 &\left.-\frac{\eta_{\alpha,\beta}\eta_{\alpha,-\beta}\cg}{\omega_{\beta}^{2}}
 \left(\omega_{\beta}\eta_{0,\beta}\cg-\omega_{\beta}\cg \eta_{0,\beta}\right)
 -\frac{\eta_{\alpha,\beta}\cg \eta_{\alpha,-\beta}}{(\omega_{\beta}\cg)^{2}}
 \left(\omega_{\beta}\cg \eta_{0,\beta}\cg-\omega_{\beta}\cg \eta_{0,\beta}\cg\right)\right.\\
 &\left.+\frac{\eta_{\alpha,0}^{2}}{2\omega_{\alpha}^{2}}
 \left(\omega_{\alpha}\eta_{\alpha,0}\cg-\omega_{\alpha}\cg \eta_{\alpha,0}\right)
 +\frac{(\eta_{\alpha,0}\cg)^{2}}{2(\omega_{\alpha}\cg)^{2}}
 \left(\omega_{\alpha}\cg \eta_{\alpha,0}-\omega_{\alpha}\eta_{\alpha,0}\cg\right)\right.\\
 &\left.-\frac{\eta_{0,\beta}^{2}}{2\omega_{\beta}^{2}}
 \left(\omega_{\beta}\eta_{0,\beta}\cg-\omega_{\beta}\cg \eta_{0,\beta}\right)
 -\frac{(\eta_{0,\beta}\cg)^{2}}{2(\omega_{\beta}\cg)^{2}}
 \left(\omega_{\beta}\cg \eta_{0,\beta}\cg-\omega_{\beta}\cg \eta_{0,\beta}\cg\right)\right.\\
 &+\left.\left(\frac{\eta_{0,\beta}}{\omega_{\beta}}+\frac{\eta_{0,\beta}\cg}{\omega_{\beta}\cg}
 -\frac{\eta_{\alpha,0}}{\omega_{\alpha}}-\frac{\eta_{\alpha,0}\cg}{\omega_{\alpha}\cg}\right)
 \left(\eta_{\alpha,0}\cg \eta_{\alpha,0}+\eta_{0,\beta}\cg \eta_{0,\beta}
 +\eta_{\alpha,\beta}\cg \eta_{\alpha,\beta}
 +\eta_{\alpha,-\beta}\cg \eta_{\alpha,-\beta}\right)\right],
 \end{aligned}
 \label{eq:5.12}
\end{equation}
in which, for convenience, we introduced two new auxiliary constants, 
\begin{subequations}
 \label{eq:5.13}
 \begin{align}
 &\chi_{\alpha,\beta}=\frac{1}{\alpha^{2}}-\frac{1}{\beta^{2}},
 \label{eq:5.13a}\\[1.2ex]
 &\xi_{\alpha,\beta}=\frac{2}{\alpha^{2}+\beta^{2}}-\frac{1}{\alpha^{2}}-\frac{1}{\beta^{2}}.
 \label{eq:5.13b}
 \end{align}
\end{subequations}

Finally, using \eqref{eq:5.10} and \eqref{eq:5.12}, we obtain the equations of motion for 
the beatified four-wave model in the following Hamiltonian form:
\begin{equation}
 \dot{\hat\eta}=J_{\hat\eta}\cdot \frac{\del\b{H}_{\hat\eta}}{\del \hat\eta}.
 \label{eq:5.14}
\end{equation}
Due to the large number of terms, we will omit the explicit calculations of \eqref{eq:5.14}. 
However, we make two additional observations. First, we emphasize that because we applied  
the beatification procedure up to the second perturbative order, equations \eqref{eq:5.14} 
are quadratically nonlinear in $\hat\eta$, in a similar way to the dynamical system of 
\eqref{eq:3.18}. Second, because of the cubic terms in the Hamiltonian~\eqref{eq:5.5}, 
it can be shown that the equations of \eqref{eq:5.14}  do  not coincide with the direct 
truncation of  \eqref{eq:5.8}. For this reason, in order to obtain the correct beatified 
four-wave model, it is  necessary to calculate the truncated values for the Poisson matrix 
and Hamiltonian function.

% Subsection 5.3 %%%%%%%%%%%%%%%%%%%%%%%%%%%%%%%%%%%%%%%%%%%%%%%%%%%%%%%%%%%%%%%%%%%%%%%%%%%%%%%%%%%%%%%%%%%%%%%%%%%%%%%%%%%%%%%%%%%%%%%%%%
\subsection{Beatified constants of motion}
\label{ssc:5.3}

As in subsection \ref{ssc:3.4}, a direct consequence of the antisymmetry of the Poisson bracket is 
that $\b{H}_{\hat\eta}$ is a constant of motion for the system~\eqref{eq:5.14}. In addition, because 
the rank of $J_{\hat\eta}$ is six, we expect to find two  Casimir invariants. These can be obtained by 
integrating linear combinations of the null eigenvectors of $J_{\hat\eta}$; however, we anticipate that  
truncated values of the functionals $\mcl{C}^{(n)}$ are likely candidates for the Casimirs, so we proceed 
by investigating them.   

Because $\mcl{C}^{(1)}$ is trivial, we begin with $\mcl{C}^{(2)}$, equation \eqref{eq:2.6} for $n=2$. 
Inserting the transformations~\eqref{eq:3.1} and \eqref{eq:4.10} into $\mcl{C}^{(2)}$ gives 
\begin{equation}
 \mcl{C}^{(2)}[\eta]=\int_{\mcl{D}} \rmd^{2}r \left(\omega_{0}^{2}+2\vep\omega_{0}\eta\right)+O(\vep^{4}). 
 \label{eq:5.15} 
\end{equation}
Then, by using the spatial Fourier expansion of the field~$\eta$ 
and retaining only the $\hat\eta$ variables of \eqref{eq:5.9}, 
we obtain the following truncated form of $\mcl{C}^{(2)}$:
\begin{equation}
 \b{\mcl{C}}^{(2)}_{\hat\eta}=2\vep\left(\omega_{\alpha}\cg \eta_{\alpha,0}+\omega_{\alpha}\eta_{\alpha,0}\cg
 +\omega_{\beta}\cg \eta_{0,\beta}+\omega_{\beta}\eta_{0,\beta}\cg\right)\,,
 \label{eq:5.16} 
\end{equation}
where time-independent terms have been dropped. Using \eqref{eq:5.16} it is readily demonstrated that 
$J_{\hat\eta}\cdot {\del\b{\mcl{C}}^{(2)}_{\hat\eta}}/{\del \hat\eta}=0$ so, indeed, $\b{\mcl{C}}^{(2)}_{\hat\eta}$ 
is a Casimir of $J_{\hat\eta}$ and a constant of motion of our system \eqref{eq:5.14}. 

Similarly, by performing the transformations~\eqref{eq:3.1} and \eqref{eq:4.10} on equation~\eqref{eq:2.6} 
for $n=3$, we obtain 
\begin{equation}
 \mcl{C}^{(3)}[\eta]=\int_{\mcl{D}}  \rmd^{2}r \left(\omega_{0}^{3}
 +3\vep\omega_{0}^{2}\eta\right)  +O(\vep^{4}),  
 \label{eq:5.17} 
\end{equation}
\noindent which, after the decomposition and truncation operations, takes the following form:
\begin{equation}
 \b{\mcl{C}}^{(3)}_{\hat\eta}=6\vep\left(
 \omega_{\alpha}\omega_{\beta}\eta_{\alpha,\beta}\cg
 +\omega_{\alpha}\cg\omega_{\beta}\cg \eta_{\alpha,\beta}
 +\omega_{\alpha}\omega_{\beta}\cg \eta_{\alpha,-\beta}\cg
 +\omega_{\alpha}\cg\omega_{\beta}\eta_{\alpha,-\beta}\right),
 \label{eq:5.18} 
\end{equation}
where the constant terms have again been removed. As expected, $\b{\mcl{C}}^{(3)}_{\hat\eta}$ 
is also a Casimir invariant for the four-wave beatified model, since it is readily seen that 
$J_{\hat\eta}\cdot {\del\b{\mcl{C}}^{(3)}_{\hat\eta}}/{\del \hat\eta}=0$. Furthermore, note 
that the expressions~\eqref{eq:5.16} and \eqref{eq:5.18} are functionally independent since 
their gradients are not parallel, i.e., they are distinct constants of motion.

Having obtained $\b{\mcl{C}}^{(2)}_{\hat\eta}$ and $\b{\mcl{C}}^{(3)}_{\hat\eta}$, a few comments are 
in order. First, one result of beatification is that the Casimirs $\mcl{C}^{(2)}$ and $\mcl{C}^{(3)}$ 
become linear in the field $\eta$ up to order $\vep^3$.  This linearity promotes a significant 
simplification of the Fourier decomposition and subsequent truncation of these constants of motion.  
Second, as noted above, the beatified four-wave system has  one more constant of motion than the system 
of \eqref{eq:3.18} obtained by direct truncation, despite the fact that  the dimensional reductions made 
in the equations~\eqref{eq:3.10} and \eqref{eq:5.8} are completely analogous. Finally, we point out that 
the perturbative order of the transformation employed in the beatification procedure does not influence 
the number of independent Casimir invariants preserved under the truncation operation. This is  because 
the  beatified Poisson operator does not depend on the order of the approximation. However, truncation 
with retention different sets  of Fourier amplitudes would yield different numbers of Casimirs, depending 
on the dimensionality of the reduced system.

%%%%%%%%%%%%%%%%%%%%%%%%%%%%%%%%%%%%%%%%%%%%%%%%%%%%%%%%%%%%%%%%%%%%%%%%%%%%%%%%%%%%%%%%%%%%%%%%%%%%%%%%%%%%%%%%%%%%%%%%%%%%%%%%%%%%%%%%%%%

% Section6 %%%%%%%%%%%%%%%%%%%%%%%%%%%%%%%%%%%%%%%%%%%%%%%%%%%%%%%%%%%%%%%%%%%%%%%%%%%%%%%%%%%%%%%%%%%%%%%%%%%%%%%%%%%%%%%%%%%%%%%%%%%%%%%%

% Section6 %%%%%%%%%%%%%%%%%%%%%%%%%%%%%%%%%%%%%%%%%%%%%%%%%%%%%%%%%%%%%%%%%%%%%%%%%%%%%%%%%%%%%%%%%%%%%%%%%%%%%%%%%%%%%%%%%%%%%%%%%%%%%%%%
\section{Numerical results}
\label{sec:6}

In this section we present a brief numerical comparison between the direct four-wave model 
of \eqref{eq:3.17} and the Hamiltonian version of \eqref{eq:5.14}. To this end we use a convenient 
tool known as the \textit{recurrence plot}\cite{Marwan07}, but a full comparison would be beyond 
the scope of the present work.

Given a vector time series~$x(t)$, with $t\in[t_{i},t_{f}]$, its associated \textit{recurrence matrix} is defined by 
\begin{equation}
 R_{j,k}(\kappa)=\Theta(\kappa-||x(t_{j})-x(t_{k})||_{\infty}),
 \label{eq:recurrencematrix}
\end{equation}
for $j,k=1,2,\ldots,M$ and $t_{j}=t_{i}+\frac{j-1}{M-1}(t_{f}-t_{i})$. In \eqref{eq:recurrencematrix},  $\Theta(s)$ 
stands for the Heaviside step function, $\Theta(s)=1$ ($\Theta(s)=0$) for $s\geq0$ ($s<0$), and the adjustable 
parameter~$\kappa$, known as \textit{threshold distance}, defines the maximum distance between two points in the 
time series for them to be considered recurrent. As a simplifying choice in \eqref{eq:recurrencematrix}, we opted  
for the supremum norm~$||\cdot||_{\infty}$, which gives the maximum absolute value among the components of its 
argument.

The recurrence plot of a signal $x(t)$ is obtained by plotting the recurrence matrix on a $t{\times}t$ plane and, 
conventionally, using black (white) dots to denote the ones (zeros) returned by $R(\kappa)$. Especially in the case 
of high-dimensional systems, the recurrence plot proves to be a powerful visualization tool, which is able to associate 
certain graphic patterns with representative behaviors of dynamical systems.

\begin{figure}[hbt]
 \begin{center}
 \subfigure{\includegraphics[width=0.3\textwidth]{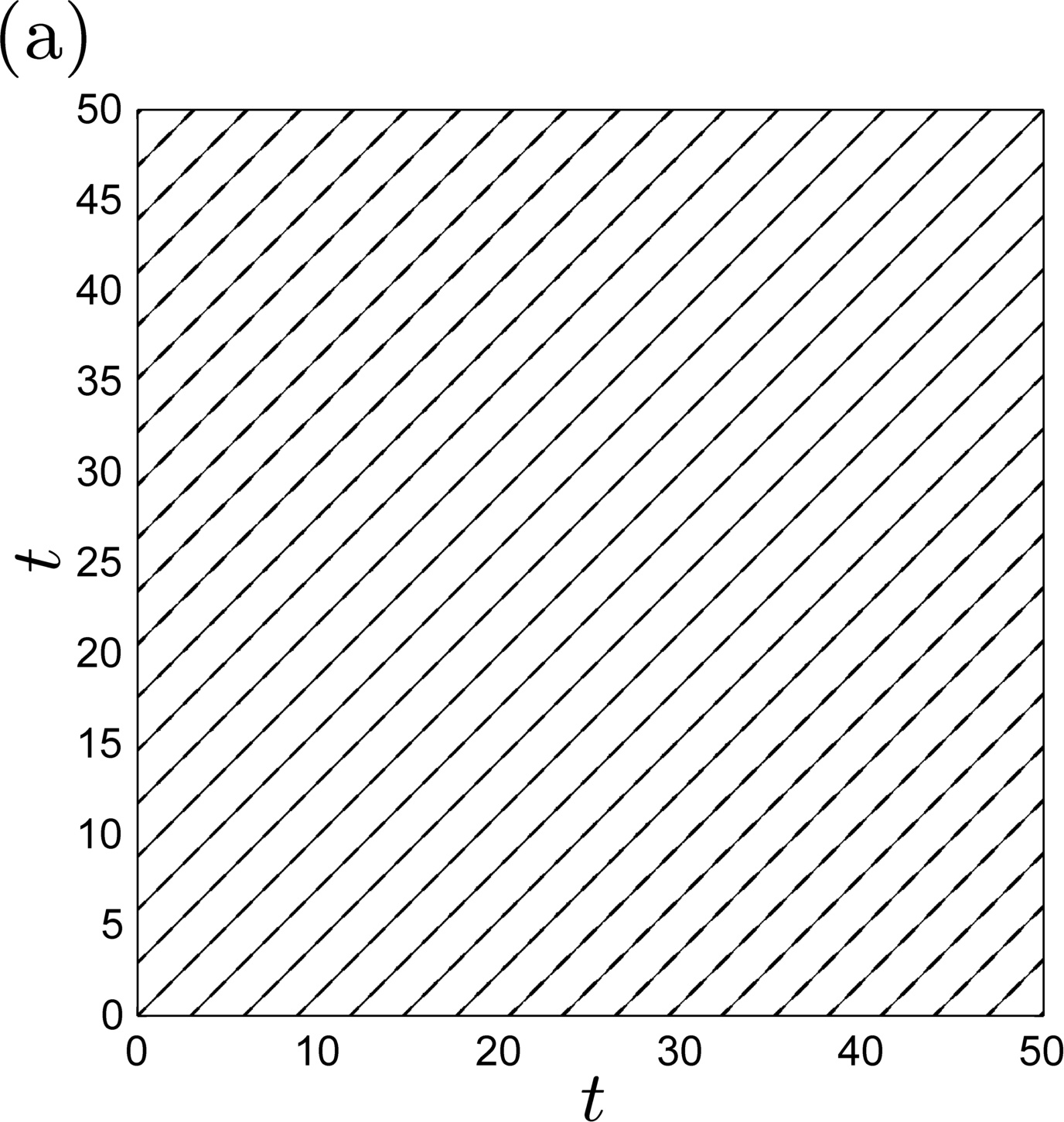}
 \label{fig1a}}
 \hspace{1.5ex} 
 \subfigure{\includegraphics[width=0.3\textwidth]{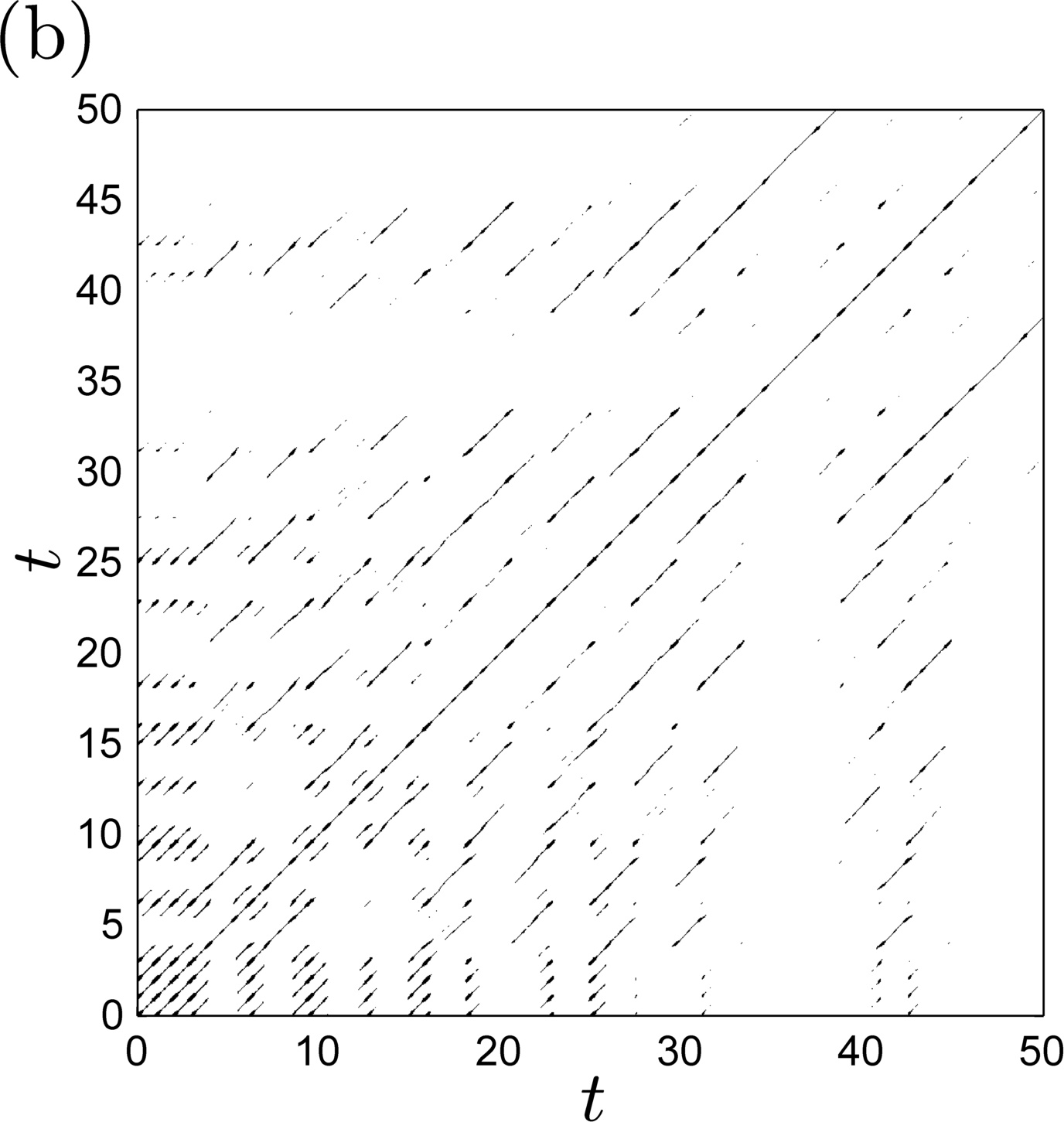}
 \label{fig1b}}
 \hspace{1.5ex}  
 \subfigure{\includegraphics[width=0.3\textwidth]{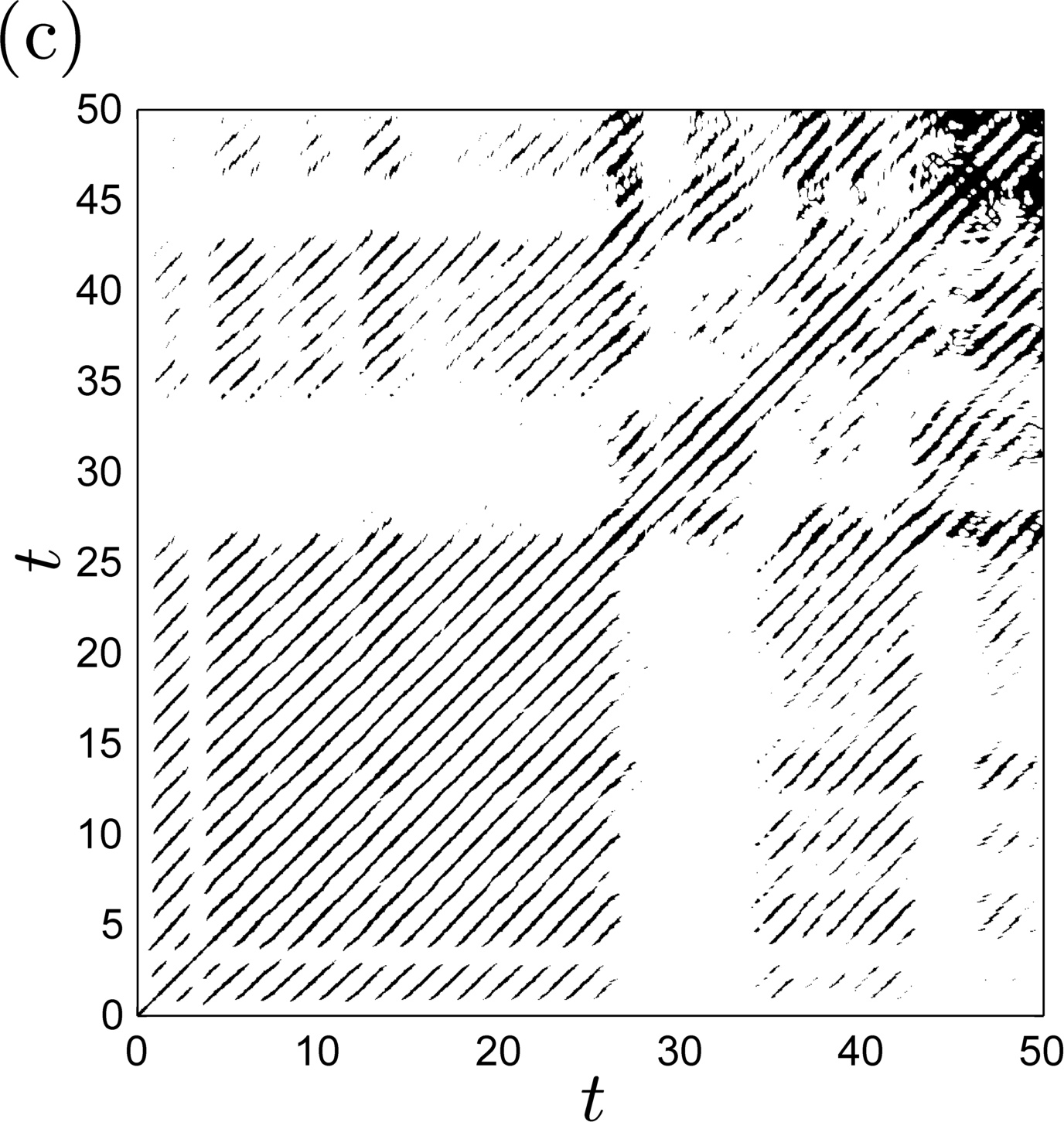}
 \label{fig1c}}
 \caption{Recurrence plot for: 
 (a) directly-truncated four-wave model of equation~\eqref{eq:3.17}, 
 (b) beatified four-wave model of equation~\eqref{eq:5.14}, and 
 (c) directly-truncated $272$-wave model obtained from equation~\eqref{eq:3.10}.}
 \label{fig1}
 \end{center}
\end{figure}

In figure~\ref{fig1}, we show the recurrence plots for three different truncation 
procedures applied to the two-dimensional Euler equation, all with parameters $t_{i}=0$, 
$t_{f}=50$, $M=1024$, and $\kappa=0.5$. For the reference state described by equation~\eqref{eq:3.2} 
we used the  parameter values $\omega_{\alpha}=3+\rmi$, $\omega_{\beta}\approx1.01-0.02\rmi$, $\alpha=7$, 
and $\beta=4$.

% Note, the exact numerical value chosen for $\omega_{\beta}$ is $(-1-2\rmi)^{0.01}$.

Figure~\ref{fig1a} displays the recurrence plot for the directly-truncated four-wave model, given 
by equation~\eqref{eq:3.17}, with initial conditions $\mu_{\alpha,0}(0)=0$, $\mu_{0,\beta}(0)=0$, 
$\mu_{\alpha,\beta}(0)=(1+\rmi){\times}10^{-3}$, and $\mu_{\alpha,-\beta}(0)=(1-\rmi){\times}10^{-3}$. 
In this particular case, for evaluating the recurrence matrix of definition~\eqref{eq:recurrencematrix}, 
we have used the time series~$x(t)=\hat{\mu}(t)$. Interestingly, for an eight-dimensional dynamical 
system with only two known constants of motion, figure~\ref{fig1a} portrays the typical pattern 
associated with periodic or quasi-periodic trajectories.

In figure~\ref{fig1b}, we present the recurrence plot for the beatified four-wave model, described 
by equation~\eqref{eq:5.14}. Accordingly, in calculating the recurrence matrix, we have used the 
time series~$x(t)=\hat{\eta}(t)$. Due to the near-identity nature of transformation~\eqref{eq:4.8}, 
for simplicity, we have employed the same initial conditions of figure~\ref{fig1a}, that is, 
$\hat{\eta}(0)=\hat{\mu}(0)$. Unlike in the case of the directly-truncated four-wave model, 
figure~\ref{fig1b} depicts the characteristic pattern of a chaotic time series, as expected 
from an arbitrarily chosen trajectory of an eight-dimensional Hamiltonian system with only 
one usual constant of motion and two Casimir invariants\footnote{This situation is equivalent 
to a Hamiltonian system with three degrees of freedom and a single constant of motion, as shown 
in appendix~\ref{app:canV}.}. 

Figure~\ref{fig1c} shows the recurrence plot for a directly truncated model obtained 
from equation~\eqref{eq:3.10} by retaining  $544$ complex amplitudes or, equivalently, 
$272$ independent spatial wave modes\footnote{The $272$-wave truncation follows the 
same reasoning described in the beginning of subsection~\ref{ssc:3.3}, where we have 
sequentially determined the most important Fourier coefficients in equation~\eqref{eq:3.10} 
for short periods of propagation.}. However, in the evaluation of the recurrence matrix, 
we have used only the time series of the eight coefficients indicated in equation~\eqref{eq:3.13}; 
that is, $x(t)=\hat{\mu}(t)$. The initial values for the four dominant waves are again the same 
as those for figure~\ref{fig1a}, while the amplitudes of the other spatial modes are initially 
zero. As seen in the figure~\ref{fig1c}, the $272$-wave model also exhibits the recurrence 
pattern associated with chaotic trajectories. 

Therefore, by considering that the $272$-wave model represents the most accurate description of the Euler's equation 
among the three dynamical systems depicted in figure~\ref{fig1}, we conclude that the beatified four-wave model presents 
a better qualitative characterization of the vorticity field's overall behavior in comparison with the directly-truncated 
four-wave model, since figure~\ref{fig1a} does not exhibit the recurrence pattern of a chaotic trajectory, as expected from 
figure~\ref{fig1c}. 

Although the recurrence patterns in the figures \ref{fig1b} and \ref{fig1c} are not exactly identical, as 
expected from such different truncation procedures, we observe that the beatified four-wave model is able 
to reproduce many features of the $272$-wave model, such as \textit{intermittency}, which is characterized 
by vertical and horizontal white stripes in the recurrence plot.

%%%%%%%%%%%%%%%%%%%%%%%%%%%%%%%%%%%%%%%%%%%%%%%%%%%%%%%%%%%%%%%%%%%%%%%%%%%%%%%%%%%%%%%%%%%%%%%%%%%%%%%%%%%%%%%%%%%%%%%%%%%%%%%%%%%%%%%%%%%

% Section7 %%%%%%%%%%%%%%%%%%%%%%%%%%%%%%%%%%%%%%%%%%%%%%%%%%%%%%%%%%%%%%%%%%%%%%%%%%%%%%%%%%%%%%%%%%%%%%%%%%%%%%%%%%%%%%%%%%%%%%%%%%%%%%%%

% Section7 %%%%%%%%%%%%%%%%%%%%%%%%%%%%%%%%%%%%%%%%%%%%%%%%%%%%%%%%%%%%%%%%%%%%%%%%%%%%%%%%%%%%%%%%%%%%%%%%%%%%%%%%%%%%%%%%%%%%%%%%%%%%%%%%
\section{Summary and conclusion}
\label{sec:7}

The main purpose of this paper is to describe a method for extracting Hamiltonian systems of finite dimension 
from a class of Hamiltonian field theories with Poisson brackets of the form of \eqref{eq:2.2}, as described 
in section \ref{sec:2}. The method was exemplified by considering a four-wave truncation of Euler's equation 
for two-dimensional vortex dynamics. In section \ref{sec:3} we described a direct method of truncation, one 
that produces equations that are energy conserving but not guaranteed to be Hamiltonian. Sections \ref{sec:4} 
and \ref{sec:5} contain the main results of the paper, the description of the method of beatification followed 
by truncation.  This  was applied to Euler's equation to produce our Hamiltonian four-wave example. Lastly, 
in section \ref{sec:6} we briefly used numerics and recurrence plots to compare our Hamiltonian four-wave model 
with the non-Hamiltonian version. 

Clearly there are many applications possible for our methodology developed here, since the class of systems 
of section \ref{sec:2} includes many models from geophysical fluid dynamics and plasma physics. Moreover, it 
is clear that the ideas pertain to more complicated Hamiltonian models such as those with more field variables, 
as are common in plasma physics modeling (see e.g.\  Ref.~\onlinecite{thiffeault}), three-dimensional magnetofluid 
models (see e.g.\ Ref.~\onlinecite{remarkable}), and sophisticated kinetic theories (see e.g.\  Ref.~\onlinecite{burby}). 
In addition, one could retain more waves in the truncation, use an alternative basis other than Fourier, and proceed to 
higher order in the beatification procedure in order to capture higher degree of nonlinearity and more complete dynamics. 
Because beatification yields a  Poisson bracket  that is independent of the dynamical variable, conventional structure 
preserving numerical methods, such as symplectic integrators, could be implemented.  

%%%%%%%%%%%%%%%%%%%%%%%%%%%%%%%%%%%%%%%%%%%%%%%%%%%%%%%%%%%%%%%%%%%%%%%%%%%%%%%%%%%%%%%%%%%%%%%%%%%%%%%%%%%%%%%%%%%%%%%%%%%%%%%%%%%%%%%%%%%

% Acknowledgements %%%%%%%%%%%%%%%%%%%%%%%%%%%%%%%%%%%%%%%%%%%%%%%%%%%%%%%%%%%%%%%%%%%%%%%%%%%%%%%%%%%%%%%%%%%%%%%%%%%%%%%%%%%%%%%%%%%%%%%%

\section*{Acknowledgements}

This work was financially supported by FAPESP under grant numbers 2011/19296-1 and 2012/20452-0 
and by CNPq under grant numbers 402163/2012-5 and 470380/2012-8. In addition, PJM received support 
from DOE contract DE--FG02--04ER--54742.

% Appendix %%%%%%%%%%%%%%%%%%%%%%%%%%%%%%%%%%%%%%%%%%%%%%%%%%%%%%%%%%%%%%%%%%%%%%%%%%%%%%%%%%%%%%%%%%%%%%%%%%%%%%%%%%%%%%%%%%%%%%%%%%%%%%%%

\appendix

% AppendixA %%%%%%%%%%%%%%%%%%%%%%%%%%%%%%%%%%%%%%%%%%%%%%%%%%%%%%%%%%%%%%%%%%%%%%%%%%%%%%%%%%%%%%%%%%%%%%%%%%%%%%%%%%%%%%%%%%%%%%%%%%%%%%%

% AppendixA %%%%%%%%%%%%%%%%%%%%%%%%%%%%%%%%%%%%%%%%%%%%%%%%%%%%%%%%%%%%%%%%%%%%%%%%%%%%%%%%%%%%%%%%%%%%%%%%%%%%%%%%%%%%%%%%%%%%%%%%%%%%%%%
\section{Beatification to second order}
\label{app:beat}

In this appendix we present the calculations leading to the beatified Poisson bracket of equation~\eqref{eq:4.14}.
The operators $\mcl{J}$, $D$, and $D^{\dg}$ defined by expressions \eqref{eq:Popt}, \eqref{eq:4.9}, and 
\eqref{eq:4.13}, respectively, satisfy various identities a few of which we will use. First, the operators $\mcl{J}$
and $D\dg$ satisfy Leibniz rules, i.e.,
\begin{subequations}
 \label{eq:LeibnizRule}
 \begin{align}
 &\mcl{J}(f)gh=g\mcl{J}(f)h+h\mcl{J}(f)g\,,
 \label{eq:LeibnizRulea}\\
 &\mcl{J}(fg)h=f\mcl{J}(g)h+g\mcl{J}(f)h\,,
 \label{eq:LeibnizRuleb}\\
 &D^{\dg}fg= gD\dg f+ fD\dg g\,,
 \label{eq:LeibnizRulec}
 \end{align}
\end{subequations}
which are true for arbitrary functions $f$, $g$, and $h$ defined on the domain~$\mcl{D}$. Second, 
the operators~$\mcl{J}$, $D$, and $D\dg$ satisfy the following interesting identity:
\begin{equation}
 Df\mcl{J}(\omega_{0})g=-\mcl{J}(f)g-\mcl{J}(\omega_{0})f{D\dg}g\,,
 \label{intID}
\end{equation}
which holds for any functions $f$ and $g$, and also any reference state~$\omega_{0}$ 
that is used in the definition of the operator~$D$.  

Inserting \eqref{eq:4.11} and its counterpart for a functional~$G$ into \eqref{eq:3.3}, then flipping the operator 
$\mathcal{S}$,  gives the following expression for the Poisson operator acting on an arbitrary function $f$:
\begin{equation}
 \begin{aligned}
 \mathcal{S}^{\dg}\mcl{J}_{\vep}(\mu)\mathcal{S}f
 =&\left[1+\vep D\mu+\frac{\vep^{2}}{2}D^{2}\mu^{2}\right]
 \left[\mcl{J}(\omega_{0})+\vep\mcl{J}(\mu)\right]
 \left[1+\vep\mu D\dg+\frac{\vep^{2}}{2}\mu^{2}(D\dg)^{2}\right]f\\
 =&\;\mcl{J}(\omega_{0})f
 +\vep\left[\mcl{J}(\mu)+D\mu\mcl{J}(\omega_{0})+\mcl{J}(\omega_{0})\mu D\dg\right]f\\
 &+\vep^{2}\left[D\mu\mcl{J}(\mu)+\mcl{J}(\mu)\mu D\dg
 +D\mu\mcl{J}(\omega_{0})\mu D\dg\right.\\
 &\left.+\frac{1}{2}D^{2}\mu^{2}\mcl{J}(\omega_{0})
 +\frac{1}{2}\mcl{J}(\omega_{0})\mu^{2}(D\dg)^{2}\right]f
 +O(\vep^{3})\,.
	\end{aligned}
 \label{eq:prebeat}
\end{equation}

Applying the identity~\eqref{intID} to the middle order $\vep$ term gives
\begin{equation}
 D\mu\mcl{J}(\omega_{0})f=-\mcl{J}(\mu)f-\mcl{J}(\omega_{0})\mu{D\dg}f\,.
 \label{eq:beps1}
\end{equation}
Thus, this term cancels the other two order~$\vep$ terms, as desired. 

Now consider the terms of order~$\vep^2$, in particular we manipulate two such terms, 
\begin{subequations}
 \label{eq:beps2}
 \begin{align}
 &\begin{aligned}
 D\mu\mcl{J}(\omega_{0})\mu{D\dg}f&=-\mcl{J}(\mu)\mu{D\dg}f
 -\mcl{J}(\omega_{0})\mu{D\dg}\mu{D\dg}f\,,
 \end{aligned}\label{eq:beps2a}\\
 &\begin{aligned}
 D^{2}\frac{\mu^{2}}{2}\mcl{J}(\omega_{0})f
 &=-\frac{1}{2}D\mcl{J}(\mu^{2})f
 -D\mcl{J}(\omega_{0})\frac{\mu^{2}}{2}{D\dg}f\\
 &=-D\mu\mcl{J}(\mu)f+\mcl{J}(\omega_{0}){D\dg}\frac{\mu^{2}}{2}{D\dg}f\,,
 \end{aligned}\label{eq:beps2b}
 \end{align}  
\end{subequations}
where all of the steps above follow from identities \eqref{eq:LeibnizRuleb} and \eqref{intID}.
Using the results of \eqref{eq:beps2a} and \eqref{eq:beps2b} all of the $\vep^2$ terms of \eqref{eq:prebeat} 
sum as follows:
\begin{equation}
 \mcl{J}(\omega_{0})\left(\frac{\mu^{2}}{2}D\dg+{D\dg}\frac{\mu^{2}}{2}-\mu{D\dg}\mu\right){D\dg}f\equiv0\,,
 \label{cancelout}
\end{equation}
as can be readily verified with the aid of identity~\eqref{eq:LeibnizRulec}. Thus, the transformation~\eqref{eq:4.8} 
flattens the Poisson operator to second order and we obtain the beatified bracket~\eqref{eq:4.14}. We observe that an infinite 
series, for which \eqref{eq:4.8} constitutes the first few terms, can be shown to flatten the bracket to all orders in a manner 
similar to, but different from, the construction of Ref.~\onlinecite{Ye91}. 

%%%%%%%%%%%%%%%%%%%%%%%%%%%%%%%%%%%%%%%%%%%%%%%%%%%%%%%%%%%%%%%%%%%%%%%%%%%%%%%%%%%%%%%%%%%%%%%%%%%%%%%%%%%%%%%%%%%%%%%%%%%%%%%%%%%%%%%%%%%

% AppendixB %%%%%%%%%%%%%%%%%%%%%%%%%%%%%%%%%%%%%%%%%%%%%%%%%%%%%%%%%%%%%%%%%%%%%%%%%%%%%%%%%%%%%%%%%%%%%%%%%%%%%%%%%%%%%%%%%%%%%%%%%%%%%%%

% AppendixB %%%%%%%%%%%%%%%%%%%%%%%%%%%%%%%%%%%%%%%%%%%%%%%%%%%%%%%%%%%%%%%%%%%%%%%%%%%%%%%%%%%%%%%%%%%%%%%%%%%%%%%%%%%%%%%%%%%%%%%%%%%%%%%
\section{Canonization}
\label{app:canV}

Beatification is the first step to canonization, by which we mean transformation to usual canonical 
variables. For the Poisson bracket of \eqref{eq:5.10} this is achieved by the following coordinate 
change:
\begin{subequations}
 \label{eq:A.1}
 \begin{align}
 &q_{1}=\rmi\sigma\rho_{\alpha}^{\frac{3}{2}}\rho_{\beta}
 \left(\omega_{\alpha}\omega_{\beta}\cg \eta_{\alpha,\beta}
 -\omega_{\alpha}\cg\omega_{\beta}\eta_{\alpha,\beta}\cg
 +\omega_{\alpha}\omega_{\beta}\eta_{\alpha,-\beta}
 -\omega_{\alpha}\cg\omega_{\beta}\cg \eta_{\alpha,-\beta}\cg\right),
 \label{eq:A.1a}\\[1.2ex]
 &q_{2}=\rmi\sigma\rho_{\alpha}\rho_{\beta}^{\frac{3}{2}}
 \left(\omega_{\alpha}\omega_{\beta}\cg \eta_{\alpha,\beta}\cg
 -\omega_{\alpha}\cg\omega_{\beta}\eta_{\alpha,\beta}
 +\omega_{\alpha}\cg\omega_{\beta}\cg \eta_{\alpha,-\beta}
 -\omega_{\alpha}\omega_{\beta}\eta_{\alpha,-\beta}\cg\right),
 \label{eq:A.1b}\\[1.2ex]
 &q_{3}=\sigma\frac{\rho_{\beta}\rho_{\alpha,\beta}^{\frac{3}{2}}}{\rho_{\alpha}}
 \left(\omega_{\beta}\cg \eta_{0,\beta}
 +\omega_{\beta}\eta_{0,\beta}\cg\right)
 -\sigma\frac{\rho_{\alpha}\rho_{\alpha,\beta}^{\frac{3}{2}}}{\rho_{\beta}}
 \left(\omega_{\alpha}\cg \eta_{\alpha,0}
 +\omega_{\alpha}\eta_{\alpha,0}\cg\right),
 \label{eq:A.1c}\\[1.2ex]
 &p_{1}=\rmi\sigma\rho_{\alpha}^{\frac{1}{2}}\rho_{\beta}
 \left(\omega_{\beta}\eta_{0,\beta}-\omega_{\beta}\cg \eta_{0,\beta}\cg\right),
 \label{eq:A.1d}\\[1.2ex]
 &p_{2}=\rmi\sigma\rho_{\alpha}\rho_{\beta}^{\frac{1}{2}}
 \left(\omega_{\alpha}\eta_{\alpha,0}-\omega_{\alpha}\cg \eta_{\alpha,0}\cg\right),
 \label{eq:A.1e}\\[1.2ex]
 &p_{3}=\sigma\rho_{\alpha}\rho_{\beta}\rho_{\alpha,\beta}^{\frac{1}{2}}
 \left(\omega_{\alpha}\omega_{\beta}\cg \eta_{\alpha,-\beta}
 +\omega_{\alpha}\cg\omega_{\beta}\eta_{\alpha,-\beta}\cg
 -\omega_{\alpha}\omega_{\beta}\eta_{\alpha,\beta}
 -\omega_{\alpha}\cg\omega_{\beta}\cg \eta_{\alpha,\beta}\cg\right),
 \label{eq:A.1f}\\[1.2ex]
 &c_{1}=\rho_{\alpha,\beta}
 \left(\omega_{\alpha}\cg \eta_{\alpha,0}
 +\omega_{\beta}\cg \eta_{0,\beta}
 +\omega_{\alpha}\eta_{\alpha,0}\cg
 +\omega_{\beta}\eta_{0,\beta}\cg\right),
 \label{eq:A.1g}\\[1.2ex]
 &c_{2}=\rho_{\alpha}\rho_{\beta}
 \left(\omega_{\alpha}\cg\omega_{\beta}\cg \eta_{\alpha,\beta}
 +\omega_{\alpha}\cg\omega_{\beta} \eta_{\alpha,-\beta}
 +\omega_{\alpha}\omega_{\beta}\eta_{\alpha,\beta}\cg
 +\omega_{\alpha}\omega_{\beta}\cg \eta_{\alpha,-\beta}\cg\right),
 \label{eq:A.1h}
 \end{align}
\end{subequations}
where $(q_{j},p_{j})$, for $j=1,2,3$, are real canonically conjugate pairs, the variables 
$c_{1}$ and $c_{2}$ are equivalent to the two Casimir invariants $\b{\mcl{C}}^{(2)}_{\hat{\eta}}$ 
and $\b{\mcl{C}}^{(3)}_{\hat{\eta}}$, and we have defined the constants $\sigma={1}/{(2\pi}{\sqrt{\alpha\beta}})$, 
%$\rho_{\alpha}=1/\sqrt{\omega_{\alpha}^{2}+(\omega_{\alpha}\cg)^{2}}$, 
$\rho_{j}=1/\sqrt{\omega_{j}^{2}+(\omega_{j}\cg)^{2}}$, for $j=\alpha,\beta$,
and $\rho_{\alpha,\beta}=1/\sqrt{\omega_{\alpha}^{2}+(\omega_{\alpha}\cg)^{2}+\omega_{\beta}^{2}+(\omega_{\beta}\cg)^{2}}$.
 
Upon writing
\begin{equation}
 \hat{\eta}_{c}=(q_{1},\,  q_{2}, \,  q_{3},\,  p_{1},\,  p_{2},\,  p_{3},\,  c_{1},\,  c_{2})\,, 
 \label{eq:A.3}
\end{equation}
the Poisson bracket for the beatified four-wave model becomes
\begin{equation}
 \{f,g\}_{\hat\eta_{c}}=\left(\frac{\del f}{\del\hat{\eta}_{c}}\right)\T\!\!\!\cdot 
 J_{\hat\eta_{c}}\cdot \left(\frac{\del g}{\del\hat{\eta}_{c}}\right)\,,
 \label{eq:A.4}
\end{equation}
with canonized Poisson matrix, 
\begin{equation}
 J_{\hat{\eta}_{c}}=\left(\begin{array}{c c}
 J_{c} & 0_{6\times 2} \\
 0_{2\times 6} & 0_{2\times 2}
 \end{array}\right),
 \label{eq:A.5}
\end{equation}
where the block $J_{c}$ is given by \eqref{eq:4.7} with $r=3$. The form of \eqref{eq:A.5} reveals  
that an ordinary three degree-of-freedom Hamiltonian system lives in the original eight dimensional 
phase space. Using the Casimir invariants, a consequence of the degeneracy of Poisson matrix, 
as coordinates separated out the superfluous dimensions, and the canonization transformation of 
this appendix put the remaining six coordinates into canonical form. 

%%%%%%%%%%%%%%%%%%%%%%%%%%%%%%%%%%%%%%%%%%%%%%%%%%%%%%%%%%%%%%%%%%%%%%%%%%%%%%%%%%%%%%%%%%%%%%%%%%%%%%%%%%%%%%%%%%%%%%%%%%%%%%%%%%%%%%%%%%%  

% Bibliografia %%%%%%%%%%%%%%%%%%%%%%%%%%%%%%%%%%%%%%%%%%%%%%%%%%%%%%%%%%%%%%%%%%%%%%%%%%%%%%%%%%%%%%%%%%%%%%%%%%%%%%%%%%%%%%%%%%%%%%%%%%%%
 
\bibliographystyle{unsrt}

\addcontentsline{toc}{section}{References}

\begin{thebibliography}{10}

\bibitem{Arakawa66}
A.~Arakawa.
\newblock {\em J. Comp. Phys.}, 1:119--143, 1966.

\bibitem{Arakawa81}
A.~Arakawa and V.~R. Lamb.
\newblock {\em Mon. Weather Rev.}, 109:18--36, 1981.

\bibitem{Salmon89}
R.~Salmon and L.~D. Talley.
\newblock {\em J. Comp. Phys.}, 83:247--259, 1989.

\bibitem{Cockburn91}
B.~Cockburn and C.-W. Shu.
\newblock {\em Math. Model. Numer. Anal.}, 25:337--361, 1991.

\bibitem{Cheng13}
Y.~Cheng, I.~M. Gamba, and P.~J. Morrison.
\newblock {\em J. Sci. Comput.}, 56:319--349, 2013.

\bibitem{Morrison80}
P.~J. Morrison and J.~M. Greene.
\newblock {\em Phys. Rev. Lett.}, 45:790--794, 1980.

\bibitem{Morrison98}
P.~J. Morrison.
\newblock {\em Rev. Mod. Phys.}, 70:467--521, 1998.

\bibitem{WW77}
J.~Weiland and H.~Wilhelmsson.
\newblock {\em Coherent Nonlinear Interaction of Waves in Plasmas}.
\newblock Pergamon Press, New York, 1977.

\bibitem{Wersinger80}
J.-M. Wersinger, J.~M. Finn, and E.~Ott.
\newblock {\em Phys. Rev. Lett.}, 44:453--456, 1980.

\bibitem{Lopes96}
S.~R. Lopes and A.~C.-L. Chian.
\newblock {\em Phys. Rev. E}, 54:170--174, 1996.

\bibitem{Sugihara68}
R.~Sugihara.
\newblock {\em Phys. Fluids}, 11:178--184, 1968.

\bibitem{Karplyuk73}
K.~S. Karplyuk, V.~N. Oraevskii, and V.~P. Pavlenko.
\newblock {\em Plasma Physics}, 15:113--124, 1973.

\bibitem{Turner80}
J.~G. Turner.
\newblock {\em Physica Scrip.}, 21:185--190, 1980.

\bibitem{Verheest82}
F.~Verheest.
\newblock {\em J. Phys. A}, 15:1041--1050, 1982.

\bibitem{Romeiras83}
F.~J. Romeiras.
\newblock {\em Phys. Lett. A}, 93:227--229, 1983.

\bibitem{Chian96}
A.~C.-L. Chian, S.~R. Lopes, and J.~R. Abalde.
\newblock {\em Physica D}, 99:269--275, 1996.

\bibitem{Pakter97}
R.~Pakter, S.~R. Lopes, and R.~L. Viana.
\newblock {\em Physica D}, 110:277--288, 1997.

\bibitem{Charney71}
J.~G. Charney.
\newblock {\em J. Atmos. Sci.}, 28:1087--1095, 1971.

\bibitem{Hasegawa78}
A.~Hasegawa and K.~Mima.
\newblock {\em Phys. Fluids}, 21:87--92, 1978.

\bibitem{Connaughton10}
C.~P. Connaughton, B.~T. Nadiga, S.V. Nazarenko, and B.~E. Quinn.
\newblock {\em J. Fluid Mech.}, 654:207--231, 2010.

\bibitem{LashmoreDavies01}
C.~N. Lashmore-Davies, D.~R. McCarthy, and A.~Thyagaraja.
\newblock {\em Phys. Plasmas}, 8:5121--5133, 2001.

\bibitem{LashmoreDavies05}
C.~N. Lashmore-Davies, A.~Thyagaraja, and D.~R. McCarthy.
\newblock {\em Phys. Plasmas}, 12:122304, 2005.

\bibitem{Kolesnikov05}
R.~A. Kolesnikov and J.~A. Krommes.
\newblock {\em Phys. Rev. Lett.}, 94:235002, 2005.

\bibitem{Dewar07}
J.P. Denier and J.S. Frederiksen, editors.
\newblock volume~6 of {\em World Scientific Lecture Notes in Complex Systems}.
  World Scientific, 2007.

\bibitem{Morrison05}
P.~J. Morrison.
\newblock {\em Phys. Plasmas}, 12:058102, 2005.

\bibitem{Tassi08}
E.~Tassi, P.~J. Morrison, F.~L. Waelbroeck, and D.~Grasso.
\newblock {\em Plasma Phys. Control. Fusion}, 50:085014, 2008.

\bibitem{Kueny95a}
C.~S. Kueny and P.~J. Morrison.
\newblock {\em Phys. Plasmas}, 2:1926--1940, 1995.

\bibitem{Kueny95b}
C.~S. Kueny and P.~J. Morrison.
\newblock {\em Phys. Plasmas}, 2:4149--4160, 1995.

\bibitem{Zakharov09}
V.~E. Zakharov and L.~A. Ostrovsky.
\newblock {\em Physica D}, 238:540548, 2009.

\bibitem{Rizzato15}
P.~Iorra, S.~Marini, E.~Peter, R.~Pakter, and F.B. Rizzato.
\newblock {\em Physica A}, 436:686--693, 2015.

\bibitem{Morrison81a}
P.~J. Morrison.
\newblock {H}amiltonian field description of the two-dimensional vortex fluids
  and guiding center plasmas.
\newblock Technical Report PPPL--1783, Princeton University Plasma Physics
  Laboratory, Princeton, New Jersey, March 1981.

\bibitem{Morrison82b}
P.~J. Morrison.
\newblock {\em AIP Conf. Proc.}, 88:13--46, 1982.

\bibitem{MorrisonVanneste}
P.~J. Morrison and J.~Vanneste.
\newblock Weakly nonlinear dynamics in noncanonical {Hamiltonian} systems with
  applications to fluids and plasmas.
\newblock Preprint: arXiv:1512.07230v1 [physics.plasm-ph].

\bibitem{Olver84a}
P.~J. Olver.
\newblock {\em Contemp. Math}, 28:231--249, 1984.

\bibitem{Olver84b}
P.~J. Olver.
\newblock Hamiltonian and {non-Hamiltonian models} for water waves.
\newblock In P.~G. Ciarlet and M.~Roseau, editors, {\em Trends and Applications
  of Pure Mathematics to Mechanics}, pages 273--290. Springer, Berlin, 1984.

\bibitem{Vorobev88}
Y.~M. Vorob'ev and M.~V. Karasev.
\newblock {\em Funct. Anal. Appl.}, 22:1--9, 1988.

\bibitem{FloresEspinoza13}
R.~Flores-Espinoza.
\newblock The lie transform method for perturbations of contravariant
  antisymmetric tensor fields and its applications to hamiltonian dynamics.
\newblock Preprint: arXiv:1308.0307v1 [math-ph].

\bibitem{Schouten53}
J.~A. Schouten.
\newblock {\em Convegno Internazionale di Geometria Differenziale}, pages 1--7,
  1953.

\bibitem{Morrison03}
P.~J. Morrison.
\newblock {H}amiltonian description of fluid and plasma systems with continuous
  spectra.
\newblock In {\em Nonlinear Processes in Geophysical Fluid Dynamics}, pages
  53--69, Dordrecht, 2003. Kluwer.

\bibitem{Morrison80b}
P.~J. Morrison.
\newblock {\em Phys. Lett. A}, 80:383--386, 1980.

\bibitem{Tassi09}
E.~Tassi, C.~Chandre, and P.~J. Morrison.
\newblock {\em Phys. Plasmas}, 16:082301, 2009.

\bibitem{Montgomery79}
D.~Montgomery and R.~Kraichnan.
\newblock {\em Rep. Prog. Phys.}, 43:547--619, 1979.

\bibitem{Jost64}
R.~Jost.
\newblock {\em Rev. Mod. Phys.}, 36:572--579, 1964.

\bibitem{Abraham87}
R.~Abraham and J.~E. Marsden.
\newblock {\em Foundations of Mechanics}.
\newblock Addison-Wesley, Redwood City, CA., 1987.

\bibitem{Sudarshan74}
E.~Sudarshan and N.~Makunda.
\newblock {\em Classical Dynamics: A Modern Perspective}.
\newblock Wiley, New York, 1974.

\bibitem{Weinstein83}
A.~Weinstein.
\newblock {\em J. Diff. Geom.}, 18:523--557, 1983.
\newblock Erratum: ibid. 22:255, 1985.

\bibitem{Chandre14}
C.~Chandre, P.~J. Morrison, and E.~Tassi.
\newblock {\em Phys. Lett. A}, 378:956--959, 2014.

\bibitem{Morrison82a}
P.~J. Morrison and J.~M. Greene.
\newblock {\em Phys. Rev. Lett.}, 48:569, 1982.

\bibitem{Viscondi11a}
T.~F. Viscondi and M.~A.~M. de~Aguiar.
\newblock {\em J. Math. Phys.}, 52:052104, 2011.

\bibitem{Viscondi11b}
T.~F. Viscondi and M.~A.~M. de~Aguiar.
\newblock {\em J. Chem. Phys.}, 134:234105, 2011.

\bibitem{Marwan07}
N.~Marwan, M.~C. Romano, M.~Thiel, and J.~Kurths.
\newblock {\em Phys. Rep.}, 438:237--329, 2007.

\bibitem{thiffeault}
Jean-Luc Thiffeault and P.~J. Morrison.
\newblock {\em Physica D}, 136:205--244, 2000.

\bibitem{remarkable}
M.~Lingam, P.~J. Morrison, and G.~Miloshevich.
\newblock {\em Phys. Plasmas}, 22:072111, 2015.

\bibitem{burby}
J.~Burby, A.~Brizard, P.~J. Morrison, and H.~Qin.
\newblock {\em Phys. Lett. A}, 379:2073--2077, 2015.

\bibitem{Ye91}
H.~Ye, P.~J. Morrison, and J.~D. Crawford.
\newblock {\em Phys. Lett. A}, 156:96--100, 1991.

\end{thebibliography}

% Fim do documento %%%%%%%%%%%%%%%%%%%%%%%%%%%%%%%%%%%%%%%%%%%%%%%%%%%%%%%%%%%%%%%%%%%%%%%%%%%%%%%%%%%%%%%%%%%%%%%%%%%%%%%%%%%%%%%%%%%%%%%%

\end{document}